# Hierarchical porous materials made by stereolithographic printing of photo-curable emulsions

Nicole Kleger[1], Clara Minas[1], Patrick Bosshard[1], Iacopo Mattich[1,2], Kunal Masania[1,3,*], André R. Studart[1,*]

[1] Complex Materials, Department of Materials, ETH Zurich, 8093 Zurich, Switzerland

[2] Soft Materials, Department of Materials, ETH Zurich, 8093 Zurich, Switzerland

[3] Current address: Shaping Matter Lab, Faculty of Aerospace Engineering, Delft University of Technology, Kluyverweg 1, 2629 HS Delft, Netherlands

*corresponding authors: k.masania@tudelft.nl, andre.studart@mat.ethz.ch

**Keywords:** Pickering Emulsions, stereolithography, hierarchical materials, porous ceramics, bio-inspired composites

## Abstract

Porous materials are relevant for a broad range of technologies from catalysis and filtration, to tissue engineering and lightweight structures. Controlling the porosity of these materials over multiple length scales often leads to enticing new functionalities and higher efficiency but has been limited by manufacturing challenges and the poor understanding of the properties of hierarchical structures. Here, we report an experimental platform for the design and manufacturing of hierarchical porous materials via the stereolithographic printing of stable photo-curable Pickering emulsions. In the printing process, the micron-sized droplets of the emulsified resins work as soft templates for the incorporation of microscale porosity within sequentially photo-polymerized layers. The light patterns used to polymerize each layer on the building stage further generate controlled pores with bespoke three-dimensional geometries at the millimetre scale. Using this combined fabrication approach, we create architectured lattices with mechanical properties tuneable over several orders of magnitude and large complex-shaped inorganic objects with unprecedented porous designs.

Hierarchical materials with pores at multiple length scales are attractive for catalysis, filtration, energy conversion and lightweight structural applications, because they feature a combination of properties that would be mutually exclusive in their conventional porous counterparts.[1] As an example, the high permeability and large accessible surface area that can be found in hierarchical porous materials are fundamental for the efficiency of filters and catalytic systems. Biological filtration systems, such as marine sponges, display hierarchical structures that combine highly permeable large channels for maximum seawater flow with small pores of high specific surface area for the uptake of large quantities of nutrients. This allows marine sponges to maximize filtration efficiency at minimum energetic costs.[2] High mechanical stiffness and low weight is another example of conflicting properties that are often observed in biological materials, such as bone and wood. In this case, hierarchical porosity provides a means to concentrate material only in regions subjected to high mechanical stresses, thus enabling weight reduction with minimum loss in the load-bearing capacity of the structure.[3, 4, 5, 6] Despite the inspiring examples from nature, the controlled manufacturing of hierarchical porous materials and the limited understanding of the structure-property relationships underlying their improved performance remains a challenge that has hampered the technological exploitation of this concept.

3D printing of self-assembling inks or resins have recently been shown to offer an effective approach to manufacture hierarchical materials with controlled multiscale porosity.[7, 8] The idea of this approach is to utilize the top-down printed patterns to define porosity at coarser length scales, while harnessing bottom-up self-assembly processes inside the feedstock material to generate fine pores at length scales below the printing resolution. This concept has been exploited to create hierarchical porous materials through direct ink writing (DIW) of emulsions and foams [9, 10, 11, 12, 13] and more recently via digital light processing (DLP) stereolithographic printing of phase-separating resins.[7] In addition to self-assembling resins, two-photon lithography approaches have also been applied for the fabrication of small-scale structures in the form of thin-walled metallic or ceramic micro- and nanolattices for fundamental mechanical studies. [8 14, 15, 16, 17] While a remarkable level of control has been achieved over the porosity of the printed multiscale structures, further development is needed to broaden the chemical compositions, macroscopic dimensions, porosity level and pore size range that can be covered by these manufacturing technologies. In particular, the stereolithographic printing of emulsions is a promising route to create tuneable microstructures[18, 19], but has not yet been exploited to manufacture hierarchical porous materials that entail high resolution, complex geometry and controlled pore sizes at multiple length scales.

Here, we design and study photo-curable stable emulsions that can be used as feedstock for the stereolithographic printing of hierarchical porous materials featuring complex geometries and controlled porosity at two distinct length scales. A key aspect of the proposed manufacturing process is the utilization of resins consisting of well-established Pickering emulsions that are stable against shear but also readily flow under low stresses. Exposing these stable emulsions to pre-defined illumination patterns using widely available desktop stereolithographic processes allows us to achieve multiscale structural control and print hierarchical architectured lattices with tuneable mechanical properties. The mechanical properties of such lattices are characterized and analysed using analytical models to gain fundamental understanding on the role of hierarchy on the deformation behaviour of bending- and stretching-dominated porous structures. Finally, we study how the printed composites can be further processed at high temperatures to yield hierarchical porous ceramics with complex geometries and bespoke porous architectures.

The manufacturing of hierarchical porous materials via DLP 3D printing of photo-curable emulsions relies on two important mechanisms for pore formation at distinct length scales. At the micrometre scale, pores are generated from the templating droplets present in the photo-curable emulsion (Figure 1a). Such droplets consist of an unreactive phase that can be removed through simple drying of the photo-polymerized part, thus creating micropores in the printed structure. To enable polymerization of the structure during the printing process, a reactive monomer mixture is used as continuous phase of the emulsion. At the millimetre scale, porosity is created by the light pattern projected onto the emulsion during polymerization of every individual layer (Figure 1b). Each polymerized layer corresponds to a slice of the 3D object generated by computer-aided design (CAD) in virtual space. The design freedom offered by the virtual tool allows for the creation of a wide range of cellular architectures from periodic lattices to topologically optimized or bio-inspired structures.

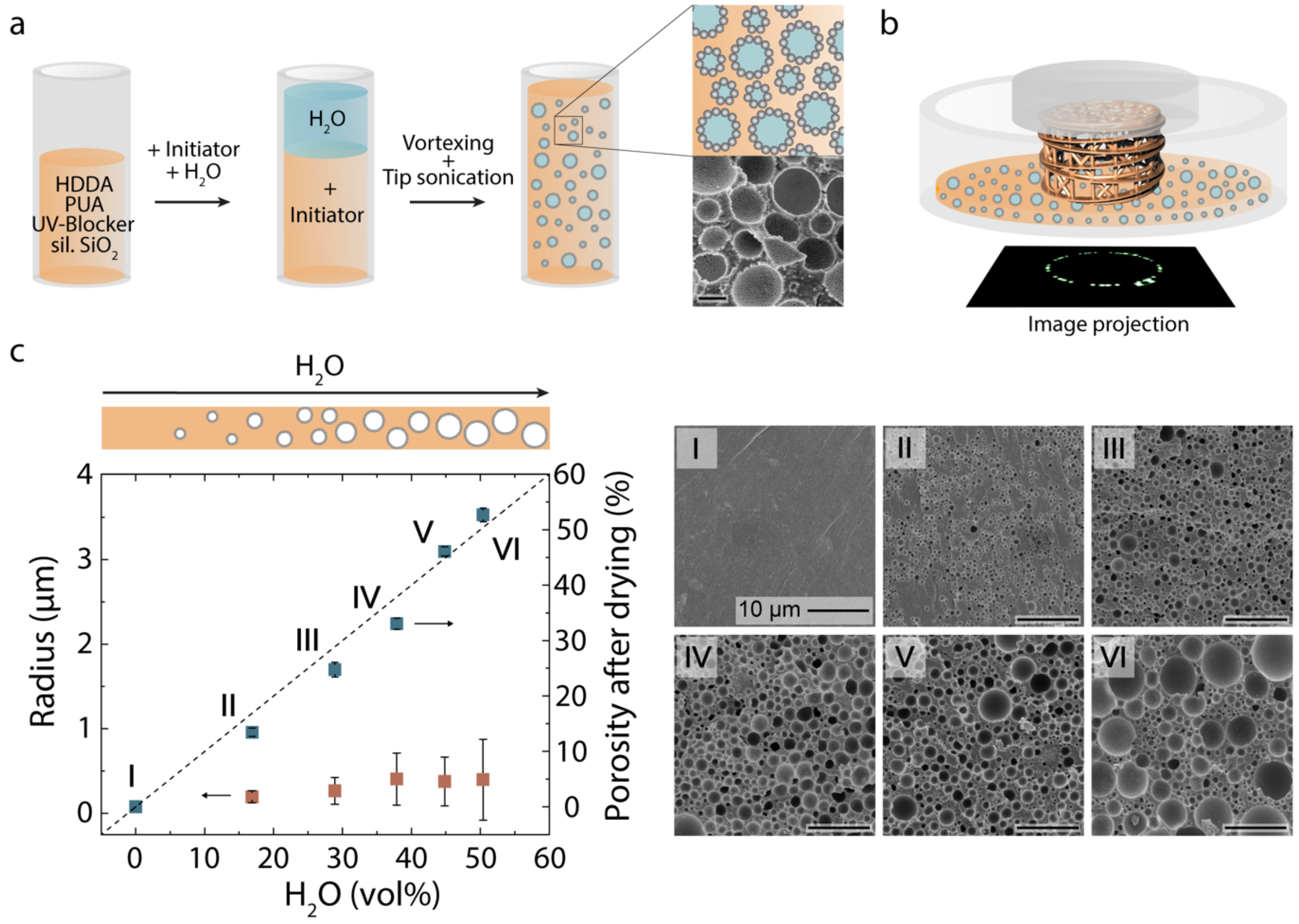

**Figure 1.** Particle-stabilized emulsions used as feedstock resin for stereolithographic printing. (a) Chemicals and processing steps used for the preparation of photo-curable water-in-oil emulsions stabilized by colloidal particles. Silanized $SiO_2$ nanoparticles are used to form a protective coating around the water droplets, whereas monomers (HDDA and PUA), a UV-blocker and a photo-initiator form a continuous oil phase that enables light-induced polymerization during printing. The scanning electron microscopy (SEM) image depicts the micropores generated upon polymerization and drying of the resin. Scale bar: 1 µm. (b) Schematics of the stereolithographic DLP process employed to print complex-shaped objects with well-defined pores at the millimetre scale. (c) Influence of the water content on the porosity (blue) and pore sizes (orange) of the structures obtained after photo-polymerization and drying. The dashed line indicates the porosity expected if all the water phase is converted into porosity. The SEM images display the microstructure of the polymerized emulsions after drying. Scale bar in (c): 10 µm.

To benefit from the design versatility provided by stereolithographic printing it is essential to formulate photo-curable emulsions that are sufficiently stable to prevent the coalescence of droplets under the shear stresses imposed during the process and that also fulfil the rheological requirements of conventional DLP resins. To prevent droplet coalescence while maintaining sufficiently low viscosity, we used Pickering emulsions in which modified silica particles adsorb at the water-oil interfaces to form a protective armour against destabilizing coalescence and coarsening events. Such silica particles also reinforce the hierarchical structure of the printed object. In contrast to emulsions used for direct ink writing, an emulsion formulation for DLP should not be stabilized by a percolating network of particles in the continuous phase, since this would result in a high yield stress below which the resin does not flow. Therefore, Pickering emulsions with a low yield stress are required for DLP printing. This was achieved through the partial silanization of 100 nm-sized $SiO_2$ nanoparticles with 3-(trimethoxysilyl)propyl methacrylate (TMS-PMA) and dispersion of the modified particles in the monomer-rich reactive phase. The concentration of surface modifier, the silica content and the monomer mixture are tuned to make the particles only partially wetted by the monomer mixture, thus promoting their self-assembly on the surface of the water droplets used as dispersed phase.

Photo-curable emulsions that resist coalescence and ripening under shearing conditions were successfully prepared using the proposed Pickering stabilization approach (Figure 1). This is demonstrated using a continuous phase comprising 1,6-hexanediol diacrylate (HDDA) and polyurethane acrylate (PUA) as monomers and Sudan I as UV-blocker that ensures localized curing during printing. A photo-initiator (Irgacure 819) is added to this mixture just before emulsification to allow for light-induced polymerization of the emulsion during printing. The emulsion is created by incorporating water into the reactive mixture using a vortexer and a tip sonicator in order to decrease the droplet size to below 10 µm. Such small droplet size is important to prevent droplet deformation and coalescence in the 50 µm gap used in the printing process. Photopolymerization of the resulting emulsion followed by drying at 60 °C leads to mechanically stable porous structures featuring pores in the low micrometre range (Figure 1a). Scanning electron microscopy (SEM) of the dried structures indicates that the modified silica particles are located predominantly on the walls of the pores. This confirms that the surface modification procedure promotes extensive adsorption of particles at the water-oil interface, resulting in a protective armour that is preserved in the printed structure after polymerization of the continuous phase.

The porosity and pore size of the printed structures can be deliberately tuned by changing the composition of the initial photo-curable emulsion. To explore the compositional design space available, we evaluate the effect of the water content, monomer ratio of the emulsion and silica concentration on the microstructure of the porous materials obtained after polymerization and drying (Figures 1c and Supplementary Figures 1 and 2). The porosity of the dried 3D-printed structures was found to be comparable to the water content initially present in the water-in-oil Pickering emulsion, indicating that the water phase does not evaporate during the printing process (Figure 1c). Pickering emulsions may exhibit catastrophic or transitional phase inversion from water-in-oil to oil-in-water if the concentration of dispersed aqueous phase or the hydrophilicity of the particles, respectively, are increased beyond a certain threshold.[20, 21] Such phase inversion is undesired in our formulation because it prevents the formation of an emulsion with a polymerizable continuous phase. For the surface modified particles and a PUA content of 10 wt% used in this experimental series, catastrophic phase inversion are observed for water contents above 50 vol%. This observation is qualitatively in line with previous work [22] and reflects the fact that the coverage of droplets by particles reduces significantly above 50 vol% water for this silica concentration. Therefore, an upper limit exists for the microscale porosity in the printed structures with silica nanoparticle contents of 20 wt% (Figure 1c).

Independent of the porosity level, average pore sizes around 0.5-2 µm were consistently achieved for all the printed structures. The small droplet sizes used to template such pores allows for the photo-polymerization of layers as thin as 50 µm without affecting the emulsion stability. Interestingly, emulsions with higher water content exhibit a broader pore size distribution, including large pores that are readily visible in scanning electron microscopy (SEM) micrographs of the printed structures. We attribute this effect to the limited availability of silica particles, which results in partial coalescence of the droplets. The pore size distribution of the printed structures can also be changed by varying the concentration of PUA relative to the other monomer (HDDA) in the emulsion (Supplementary Figure 1). Finally, an increasing weight fraction of $SiO_2$ particles results in a more narrow pore size distribution (Supplementary Figure 2), which underlines the impact of particle availability for interface stabilization.

In order to better understand the effect of the PUA and water contents on the printability, rheological measurements of the yield stress ($\tau_y$) and high frequency viscosity ($\eta_\infty$) of our formulations were performed (Supplementary Figure 3). The yield stresses in the range of 0.2 – 32 Pa measured for our resins are lower than those typically found in DIW inks. [9, 12, 13] Importantly, the addition of PUA monomer lowers the yield stress of formulations with high water content compared to pure HDDA monomer emulsions. In the absence of PUA, the formation of larger droplet sizes increases the amount of non-adsorbed particles available to build a network of higher yield stress. The incorporation of PUA reduces the droplet size and increases the total droplet surface area, thus reducing the amount of non-adsorbed, network-forming particles between droplets in the emulsion. With $\tau_y$ values below 15 Pa, the emulsions with PUA reach a yield stress level that is very close to the range 6-10 Pa of other particle-laden suspensions printed by DLP. [23] Moreover, the high-frequency viscosity below 20 Pa.s of our emulsions is comparable to the typical values (10-60 Pa.s) reported in the literature for DLP printing of suspensions. [24, 25] These rheological properties allowed for the successful printing of Pickering emulsions into complex-shaped structures with pore sizes at multiple length scales.

While self-assembly processes control the porosity at the microscale, pores in the millimetre range are created through the controlled photo-polymerization of specific regions of each emulsion layer during the printing procedure. To generate millimetre pores with high fidelity in bespoke geometries it is essential to control the photo-polymerization process within the layer (x-y plane) and along the build direction (z-axis). The main compositional and processing parameters that affect the fidelity of the printed parts are the concentrations of initiator and UV-blocker in the photo-curable emulsion, as well as the illumination conditions employed during printing, such as light dose and exposure time.

To identify emulsion formulations and printing conditions that lead to high-fidelity porous structures, we first analyse the effect of compositional and processing parameters on the polymerization behaviour of the reactive layer. By applying Lambert-Beer's law to quantify the decay in light intensity along the build direction, the effect of the initiator and UV-blocker concentrations on the measured thickness of the polymerized layer can be predicted reasonably well, allowing us to gain control over the photo-induced polymerization process (Figure 2a,b). Using emulsion formulations with optimal concentrations of initiator (0.5 wt% with respect to monomer) and UV-blocker (0.075 wt% with respect to monomer), we identified the illumination conditions leading to high-fidelity printed structures with negative and positive feature sizes of 0.1mm and 0.05mm, respectively (Supplementary Figure 4).

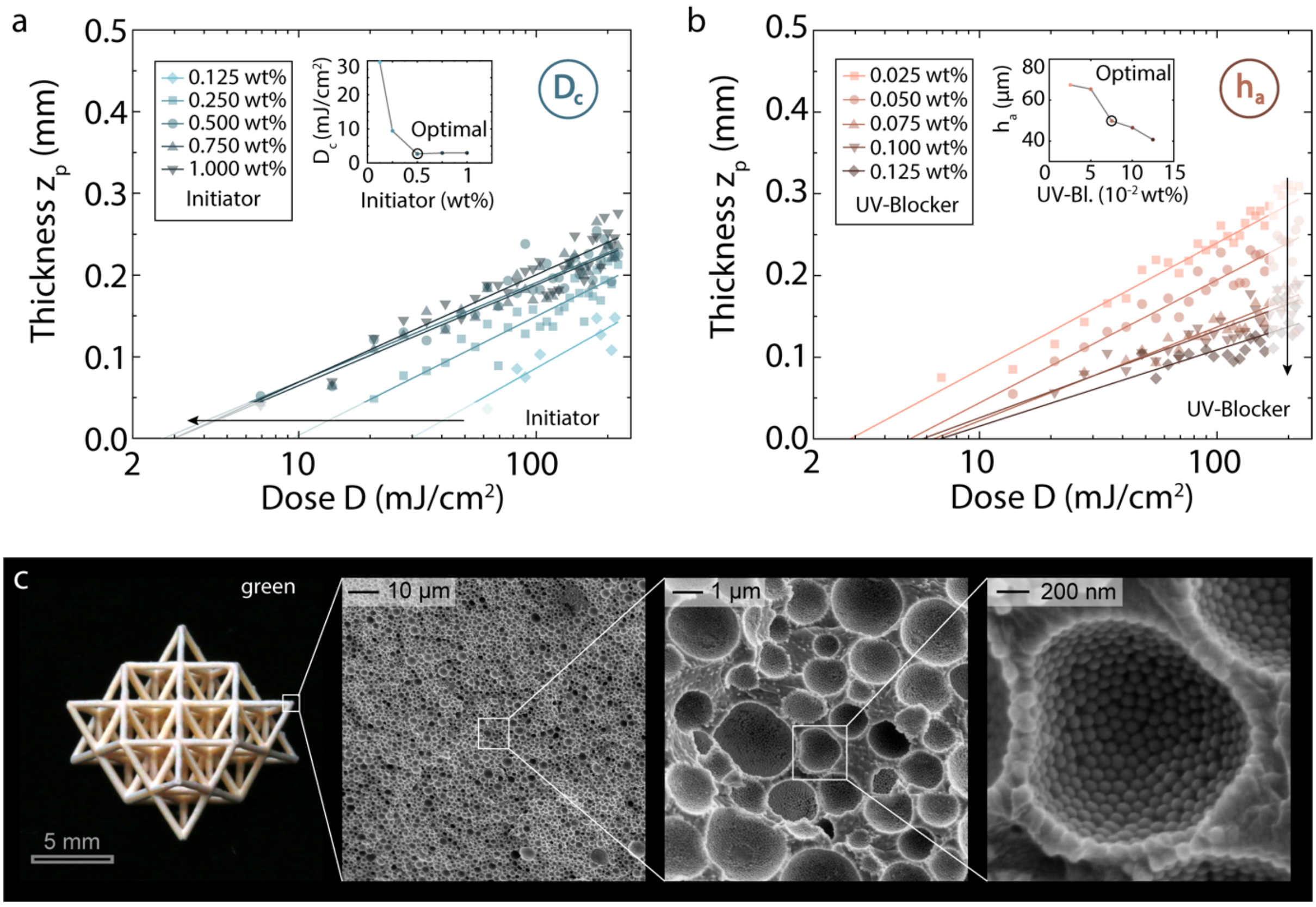

**Figure 2.** Photo-polymerization behaviour of photo-curable emulsions and DLP printing of hierarchical lattices. (a,b) The effect of the light dose (*D*) on the thickness of the photo-polymerized layer ($z_p$) for emulsions containing (a) initiator concentrations between 0.125 and 1 wt% (0.075 wt% UV blocker) with respect to the monomer, or (b) UV blocker contents varying from 0.025 to 0.125 wt% (0.5 wt% initiator) with respect to the monomer. The dependence of the critical dose ($D_c$) and of the penetration depth ($h_a$) on the concentration of initiator and UV blocker, respectively, is shown in the insets. (c) Hierarchical octet lattice prepared by DLP printing of an optimum emulsion formulation using a light intensity of 20 mW/cm$^2$ for 3 seconds per polymerization layer. The photograph reveals the open cells at the macroscale, whereas SEM images illustrate the microstructure of the lattice struts at different magnifications.

Layer-by-layer stereolithographic printing of our emulsions using optimized illumination conditions allows for the fabrication of complex composite structures with well-defined struts and pores at two distinct length scales (Figure 2c). As an illustrative example, we printed a three-dimensional octet lattice that features periodic struts down to $0.125 \pm 0.05$ mm defined by the stereolithographic process and micron-sized pores generated from the water droplets from the initial emulsion. SEM images of the structure reveal that the struts of the lattice contain a high density of pores with size of $0.50 \pm 0.26$ µm. At the selected water fraction of 45 vol%, the droplets in the templating emulsion are spherical and loosely packed. Because the emulsion templates are stabilized by interfacially adsorbed particles, an ordered array of monodisperse silica nanoparticles is formed at the internal walls of the pores. The possibility to tune the geometry of the lattice, the strut thickness and the pore sizes in a controlled manner makes these hierarchical materials attractive for fundamental and application-oriented studies.

To investigate structure-property relationships of these unique hierarchical structures, we designed, printed and characterized the mechanical properties of two exemplary 3D lattices displaying tunable porosity at the macro- and microscale (Figure 3). In this experimental study, octet and tetradecahedral (Kelvin) lattices were used as examples of architectured lattices with elastic properties dominated by, respectively, stretching and bending of the load-bearing solid elements. Variation of the strut thickness and the incorporation of pores within the struts were the strategies used at the macro- and microscales to change the total relative density of the lattice over a broad range from 0.015 to 0.500.

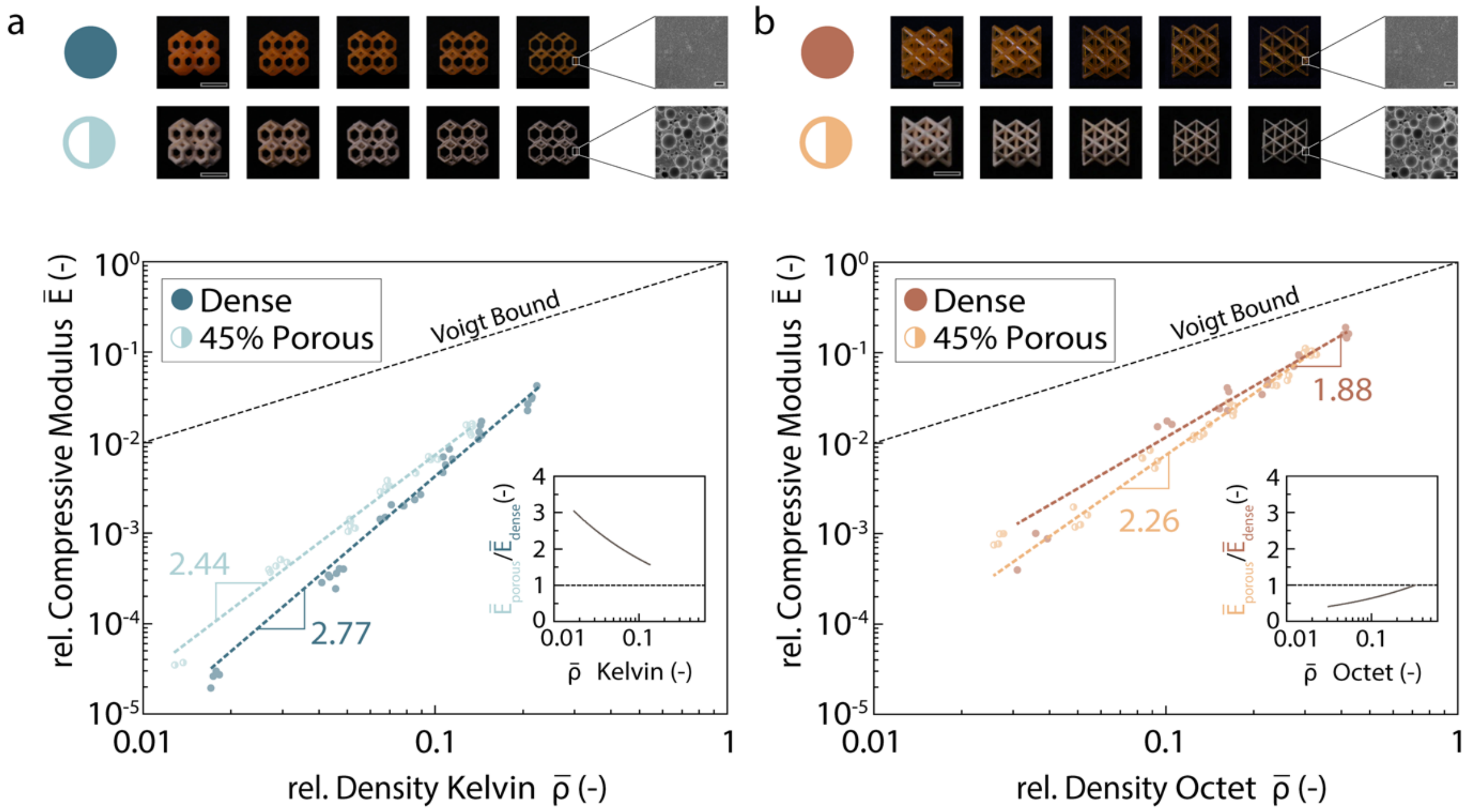

**Figure 3.** Mechanical properties of Kelvin and octet hierarchical lattices with varying relative densities. (a,b) Relative compressive modulus ($\bar{E}$) of (a) Kelvin and (b) octet lattices with struts that are either dense or 45% porous. In these samples the relative density is changed by varying the strut thickness. The insets show the ratio between the relative compressive modulus for microporous and dense Kelvin lattices ($\bar{E}_{porous}/\bar{E}_{dense}$) at varying relative densities of the overall structure ($\bar{\rho}$). Scale bars: 5 mm in macroscopic images of lattices (Kelvin and octet); 10 µm in SEM close-ups of dried emulsions.

The role of pores at distinct length scales on the mechanical properties of the printed hierarchical structures was assessed by measuring the compressive strength and elastic modulus of lattices with different relative densities and comparing the obtained experimental data with analytical scaling models available in the literature. To set a reference, we first consider Kelvin and octet lattices with dense struts of variable thickness (Figure 3). Our experimental results indicate that the octet lattice exhibit significantly higher relative elastic modulus ($E_r$) and a weaker dependence of $E_r$ on the relative density compared to the Kelvin samples. In line with previous works [26, 27], the higher stiffness of the octet specimens reflects the fact that the mechanical properties of such lattices are dominated by stretching of the struts. Struts subjected to stretching offer stronger resistance against the imposed mechanical load in comparison to the bending-prone struts of the Kelvin lattices.

To analyse the dependence of the normalized elastic modulus ($E_r$) on the relative density of the lattice ($\rho_r$), we use the simple scaling relation[6, 28, 29, 30]: $E_r = E/E_s = C\rho_r^a$, where $E$ is the absolute elastic modulus of the lattice, $E_s$ is the elastic modulus of the solid phase, $a$ is the power law exponent and $C$ is a constant. Fitting this equation to our experimental data leads to power exponents ($a$) of 2.77 and 1.88 for the Kelvin and octet lattices, respectively. These values fall within the range obtained in previous experiments, but differ from the exponents expected from analytical models and simulations.[26] The exponent measured for the Kelvin lattice ($a = 2.77$) is higher than the value of 2 predicted for bending-dominated foam structures.[29] Likewise, the elastic modulus of the octet lattice decays with the volume fraction of voids at a faster rate ($a = 1.88$) compared to the theoretical limit for stretching-dominated structures (Voigt bound, $a = 1$). Earlier work suggests that such disagreement may arise from the rotation of the nodes of non-slender beam lattices under compression.[26, 31]

After the analysis of such reference structures, we study the effect of hierarchical porosity on the mechanical properties of the lattices. To this end, a fixed volume fraction of micropores of 45 vol% is incorporated into the struts of Kelvin and octet lattices featuring variable strut thicknesses (Figure 3a,b). For bending-dominated Kelvin lattices, the presence of microporosity in the struts leads to hierarchical structures that are up to 3 times stiffer than the reference specimens (Figure 3a, inset). In the $\bar{E} - \bar{\rho}$ plot, this is manifested by a shift of the elastic modulus data towards lower relative densities and by a weaker dependence of $\bar{E}$ on $\bar{\rho}$ in comparison to the reference lattice. The stiffening effect arising from the hierarchical design replicates the strategy found in biological materials and macroscopic engineering structures to produce lightweight architectures with minimum loss in mechanical properties.[32] For load-bearing elements subjected to bending, such high mechanical efficiency is associated with the removal of solid phase from the core of the bending element where the applied stresses are lower compared to the outer shell.[27]

Simplified theoretical models based on lattices with high-aspect-ratio struts predict that the incorporation of pores at the finer scale does not change the scaling exponent.[32] The lower exponent measured for the hierarchical lattice ($a = 2.44$) suggests that mixed deformation modes also play a role on the mechanical response of structures containing pores at multiple length scales. Since lattices with high-aspect ratio struts are more prone to bending, the hierarchical design is most beneficial in lattices with long struts and ultra-low relative densities. The probable effect of mixed deformation modes and the challenge of manufacturing the idealized structures considered in simple analytical models [6] prevents a direct comparison of our experimental results with previous theoretical descriptions. In fact, simplified analytical models do not predict the experimentally observed dependence of the relative elastic modulus on the hierarchy of the porous structure. This calls for the development of new simulation and analytical approaches for the prediction of the mechanical properties of hierarchical porous materials.

In contrast to the Kelvin architecture, mechanical analysis of octet lattices indicates that the presence of a bending-dominated foam within the struts increases the compliance of the stretching-dominated structure at the macroscale (Figure 3b). Indeed, the hierarchical octet structures are up to 2 times more compliant and show a stronger dependence of $\bar{E}$ on $\bar{\rho}$ compared to reference specimens (Figure 3b, inset). The higher power exponent ($a = 2.26$) obtained for the hierarchical structures might be related to enhanced bending effects predicted by theoretical modelling and simulations [27, 33], but thus far not experimentally confirmed.

Another approach to change the relative density of our hierarchical lattices consists in the incorporation of an increasing volume fraction of micropores in the struts while keeping their thickness constant. The elastic modulus of hierarchical structures with tuneable strut porosity was found to display similar trends with respect to the relative density (Supplementary Figure 5). The same rationale also applies for the maximum compressive strength of our architectured lattices (Supplementary Figure 6).

Overall, the ability to independently control the strut thickness and the volume fraction of micropores within the struts allows us to cover a wide range of elastic properties using the same set of chemical building blocks. This distinguished feature of our architectured lattices is one of the key design concepts of biological hierarchical structures [3] and is a compelling strategy to tailor material properties using a narrow pallet of resources.

To enlarge the pool of possible applications and enable control over other structural features, the additively manufactured composites can also be converted into hierarchical porous materials with an entirely inorganic composition by calcining and sintering the printed structures at high temperatures. Using a hierarchical octet lattice as an example, we observe that the multiscale structure of the printed composites are highly preserved after the calcination and sintering steps, in spite of the high shrinkage associated with the removal of the organic phase and densification of the silica nanoparticles (Figure 4a). Similar to established injection moulded ceramics[34], slow removal of the organic phase is essential to prevent cracking of the structure during the heat treatment. Once this condition is fulfilled, the sintering process can be exploited to control the porosity level. Because the organic phase is completely removed, the size and morphology of the pores at the microscale are fully given by the porosity generated by the Pickering emulsion at a length scale that is well below the resolution of the printer (Supplementary Figure 7).

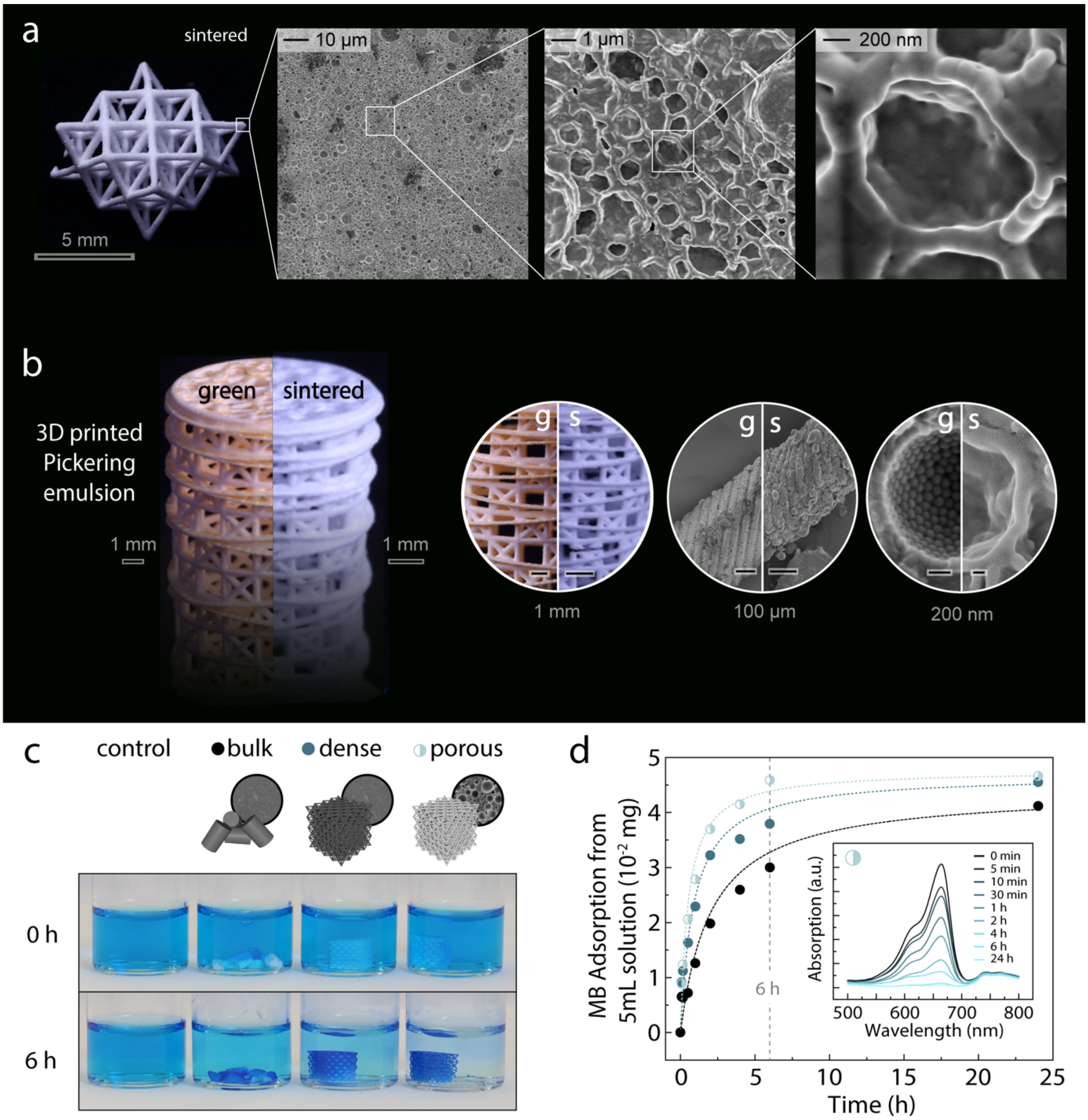

**Figure 4.** Hierarchical silica structures obtained upon heat treatment of as-printed composites. (a) Octet lattice with microporous silica struts. SEM images of the struts reveal that sintering

of the silica nanoparticles at high temperatures result in high shape retention also in the microstructure. (b) Complex hierarchical structures fabricated via 3D printing of Pickering emulsions. (c, d) Adsorption measurements of the bulk ceramic and octet lattices with dense and microporous struts using a 10 mg/L aqueous methylene blue (MB) solution. The UV-Vis measurements (d) show that the octet lattice with microporous struts shows an adsorption rate that is a factor eight faster than the bulk ceramic. The dashed lines correspond to a theoretical fitting using a pseudo-second order adsorption kinetics model. [35] The inset in (d) shows the evolution of the absorption spectra of MB aqueous solution exposed to the octet lattice with microporous strut over a period of 24 hours.

While hierarchical porous materials in the form of periodic lattices find many potential applications, the stereolithographic printing of photo-curable emulsions can also be exploited to manufacture structures with other complex multiscale architectures. For instance, materials with graded porosity along the build direction are obtained if emulsions with different water droplet fractions are used as feedstock in the layer-by-layer printing process (Supplementary Figure 8a). Using a simple cylindrical geometry to illustrate this approach, we printed a graded macroscopic object featuring a stepwise increase in local porosity from 0 to 50 vol% while keeping sufficient bonding between individual layers to prevent delamination or cracking even after calcination and sintering (Supplementary Figure 8b).

The multiscale structural control offered by our processing platform is also demonstrated by printing a 3D object that incorporates some of the design concepts of the glass skeleton of the marine sponge *Euplectella sp* (Figure 4b).[36, 37, 38] Analogous to the biological structure, the printed object exhibits a cylinder cage geometry with diagonally-reinforced walls made of a chessboard-like squared lattice at the millimetre scale, which in turn is built from micron-thick struts with a laminar design defined by the printed layers. Each individual layer of the struts is formed by an assembly of nanoparticles that are densely compacted in the biological sponge and arranged around micropores in the synthetic counterpart. Remarkably, this intricate multiscale architecture can be preserved after sintering to yield hierarchical porous materials with unparalleled level of structural control.

To illustrate a proof-of-concept potential application of hierarchical open porous ceramics, we evaluate the ability of our 3D printed porous octet structures to remove molecules from water via adsorption on the silica surface (Figure 4c,d). Because the silica surface is negatively charged in water, it can be used to adsorb positively charged pollutants. Taking positively charged methylene blue (MB) as exemplary pollutant, we measured the adsorption kinetics of MB molecules on silica octet lattices featuring microporous struts and compared it with that of octets with dense struts and with bulk silica pieces (see SI).[35] The increased surface area of the hierarchical porous octets has a marked effect on the adsorption performance, increasing by eight-fold the adsorption rate in comparison with the bulk silica ceramic. Compared to the dense octet truss structure, the hierarchical porous octet remains twice as fast in adsorbing the model pollutant, despite having half as much material present. This demonstrates a remarkable attribute of Nature’s frugal material usage through efficient hierarchical design. The hierarchical design also allows the octets with porous struts to combine the high permeability of its macropores with the high surface area provided by the microporosity. Due to the high temperature stability of the silica structures, they could be heated to 900 °C to remove the adsorbed methylene blue and recycled and reused as adsorbing sponges with the same performance as the pristine samples (Supplementary Figure 10).

In summary, hierarchical porous structures with complex macroscopic geometry and pore families at the millimetre and micrometre scales can be fabricated using particle-stabilized emulsions as feedstock resin in a desktop stereolithographic printer. Microscale porosity up to 50% and pore sizes in the range 0.5-2 μm are created by changing the volume fraction and size of droplets suspended in the reactive monomer mixture used as continuous phase of the feedstock emulsions. Light-patterning of such emulsions in a layer-by-layer fashion leads to complex-shaped pores that reach sizes down to 0.1 mm if the concentrations of initiator and light absorber in the emulsion are properly optimized. Hierarchical lattices fabricated via this approach show an intricate structure of well-defined open cells at the millimetre scale, the struts of which feature up to 50 vol% internal porosity. The mechanical properties of such lattices can be tuned within 4 orders of magnitude by changing the porosity at multiple scales and the balance between bending and stretching struts present in the structure. Heat treatment of the as-printed materials results in hierarchical porous ceramics featuring complex macroscopic geometries and tuneable pore morphology, porosity gradients, and pore sizes across multiple length scales. Combining self-assembled emulsions with the shaping freedom of stereolithographic printing is therefore an effective strategy to create hierarchical porous materials with unparalleled structural complexity and mechanical properties that are programmable over a broad range without altering the chemistry of the constituent building blocks. Such manufacturing capabilities may be harnessed to produce more efficient filters, sustainable lightweight structures and tuneable architectured materials through multiscale structural design.

## Acknowledgements

We thank the financial support from the Swiss National Science Foundation (SNSF) through Sinergia (Project number CRSII5_177178) and Consolidator grants (Project number BSCGI0_157696). This work also benefitted from the SNSF support via the National Center of Competence in Research Bio-Inspired Materials. Lorenzo Barbera and Dr. David Moore are greatly acknowledged for their help with the printing calibration and the inputs on the chemical compositions, respectively. We also thank Murielle Schreck and Madeleine Fellner for their support in the adsorption experiments and UV-Vis measurements. Furthermore, the authors wish to thank Emanuele Colucci for his support with the particle silanization.

## Author contributions

N.K., C.M., P.B., I.M., K.M. and A.R.S. developed the idea together and designed the experiments. Concerning the experimental work, N.K. and C.M. focused on the emulsion development and analysis; I.M. developed the silanization procedure and monomer composition. N.K. worked also on the mechanical analysis and sintering protocol; P.B. designed and performed the print calibration experiments and helped optimizing the resin composition for the emulsions; and K.M. programmed the CAD structures and contributed with the mechanical data analysis. All remaining experimental work was carried out together by N.K., C.M., P.B., I.M. and K.M. N.K., C.M., P.B., I.M., K.M. and A.R.S. conducted the data analysis and co-wrote the paper. All authors discussed the results of the experiments and analysis as well as their interpretation, and revised the manuscript at all stages.

## Competing interests

The authors declare no conflicting interests.

## Supporting Information

Supplementary information is available via the online version.

## Experimental Section

Emulsion preparation

$SiO_2$ nanoparticles (AngströmSphere, 100 nm diameter) were purchased from Fiber Optics Center. For particle hydrophobization, 5 g of $SiO_2$ particles were dispersed in 20 g of toluene and 0.25 mL of distilled $H_2O$ by tip sonication for 30 min at 80% power in an ice bath to prevent overheating. 1.25 g of the silanization agent 3-(trimethoxysilyl)propyl methacrylate (TMS-PMA, 98% purity, Aldrich-Fine Chemicals) and 0.12 mL of butylamine (BA, 99%, ABCR GmbH) were added to the silica suspensions and allowed to adsorb on the silica surface for 15 min during tip sonication at 80% power in the ice bath. The process was stopped by centrifugation of the particles at 4500 rpm for 10 min followed by removal of the supernatant. The hydrophobized particles were further washed three times with toluene and dried in a Petri dish at 40 °C. The monomers 1,6-hexanediol diacrylate (HDDA, > 85%, Tokyo Chemical Industry) and polyurethane acrylate (PUA, Neorad U-30W, DSM-AGI) employed for the preparation of the photo-curable resin were used as received. The UV blocker Sudan I (97%, Sigma-Aldrich) was dissolved in HDDA and mixed with PUA to adjust the light absorption properties of the resin. The silanized $SiO_2$ particles were dispersed in this resin by tip sonication for 30 min at 80% power in an ice bath. The photoinitiator phenyl bis(2,4,6-trimethylbenzoyl)-phosphine oxide (Irgacure 819, BASF) was further dispersed in the resin, followed by the addition of distilled $H_2O$ as an immiscible second phase. A primary emulsion was formed by vortexing. Thereafter, two sequential 15 min tip sonication steps at 80% power in an ice bath were performed with intermediate vortexing to allow further break-up of droplets to the size required for printing. The water content of the emulsions was set to 0 vol%, 16.9 vol%, 28.9 vol%, 37.9 vol%, 44.9 vol% and 50.4 vol%. For simplicity, these numbers are reported as rounded full integers in the manuscript.

Pore-size and porosity characterization

Scanning electron microscopy (SEM) images of porous samples were analysed using the ImageJ Software for Mac (version 1.51w, https://imagej.net/Fiji/Downloads). For each data point, the sizes of at least 700 pores were analysed. The equivalent radius $r$ was calculated from the area $A$ of a pore using the relation: $r = ( A/\pi )^{0.5}$ and is referred to as 'droplet size'. The density $\rho$ of cylindrical 3D printed samples was calculated by dividing their masses by the geometrically-measured volumes. The porosity $P$ was calculated by $P = \frac{\rho_{porous}}{\rho_{dense}}$, where $\rho_{dense} = 1.3297\ g/cm^3$ is the density of cylinders prepared from resins without water and $\rho_{porous}$ is the density of the porous sample. At least 3 cylinders were evaluated for each data point.

3D printing of Pickering emulsion resins

A commercially available DLP 3D printer (Ember, Autodesk) with the corresponding slicer software Autodesk ® Print Studio (version 1.6.5. for Mac, Autodesk) was used for 3D printing and print file generation, respectively. The printer was customized to print with small quantities of ink (< 10 ml) in a coated Petri dish. In order to successfully print with minimal adhesion of the print part to the Petri dish, the latter was lined with polydimethylsiloxane (PDMS) for emulsions or PDMS covered with a fluorinated ethylene propylene (FEP) film (McMaster Carr) for inks without water. The tray was further shielded with an elastic plastic wrap to prevent extensive evaporation of the water from the emulsion. The octet trusses, TKDH, seasponge and cylinder macrostructures were designed with SolidWorks (Dassault Systèmes). Samples for mechanical testing were printed on support poles in order to minimize defects on the structure. In order to assure good adhesion of the part to the print head, the first and the following 4 burn-in layers are illuminated for as much as 30 s and 40 s, respectively, as compared to the optimized illumination time of 3 s for the model layers used for the actual structure. After calibration using a light meter (G&R Labs Model 222), the actual illumination intensity was fixed at 20 mW/cm$^2$ (nominal intensity of 21.6 mW/cm$^2$), whereas the layer height was set to 50 $\mu$m. The UV blocker and initiator concentrations were optimized by illuminating square areas for varying durations in multiples of 0.32 s at constant intensity. The obtained thickness was measured with a micrometre. For the optimization of illumination time of the model layers, poles and holes of varying diameter were printed at 2, 3 or 4 s exposure time per layer. The resulting geometries were visualized on a Keyence Microscope and measured with ImageJ.

Mechanical analysis of hierarchical porous structures

The high shape flexibility of the printing process combined with the option to integrate microporosity within the printed parts offer a platform to fabricate and study the mechanical behaviour of hierarchical porous structures. The mechanical analysis was performed on the printed composite structures. Samples were printed with the optimized settings described above, followed by drying over night at 60 °C to eliminate the water from the pores. Finally, the samples were post-cured for 2 x 5 min with a UV lamp (Omnicure S1000, Lumen Dynamics). The mechanical properties of porous samples were measured in compression (Shimadzu AGS-X mechanical tester, Shimadzu Ltd.) with a 1kN load cell at 0.5 mm/min. Dense cylinders (no water) were tested in another mechanical testing machine (Instron 8562) with a 100kN load cell using a displacement rate of 0.5 mm/min. The compliance and strain data were corrected to account for the elasticity of the testing machine and slight deviations from parallel surfaces, respectively. The relative maximum compressive strength ($\bar{\sigma}$), elastic modulus ($\bar{E}$) and density of the samples ($\bar{\rho}$) were calculated based on the following properties of dense cylinders (0 vol% $H_2O$): $\rho_s$= 1.3297 g/cm$^3$, $E_s$ = 857.801 MPa and $\sigma_s$ = 304.081 MPa. To analyse the experimental results, the equation $Y = B * X^A$ was fitted to the absolute data using the software Package OriginLab (OriginPro 2019). To find starting parameters $A^*$ and $B^*$, the linearized equation $log(Y) = log(B^*) + A^* * log(X)$ was used. Orthogonal distance regression was applied to iteratively minimize $\chi^2$ by changing $A^*$ and $B^*$ to result in $A$ and $B$, respectively. The parameters *A* and *B* resulting from the fitting were converted to the parameters *a* and *b* that describe the relative data using the relations: $a = A$ and $b = \frac{B}{\bar{X}} * 1.3297^A$ , where $\bar{X}$ is the corresponding mean absolute value of the parameter investigated. The exponents reported in Figure 3 refer to the fitting parameter *a*.

<u>Pyrolysis and sintering of printed samples</u>

The pyrolysis process was investigated by performing thermogravimetric analysis (TGA) on a printed and dried emulsion prepared with 45 vol% $H_2O$ (Netzsch STA 449 C Jupiter ®, Netzsch-Gerätebau GmbH). The measurement was run in an $Al_2O_3$ crucible at a heating rate of 2 °C/min between 40 and 800 °C in air. From this analysis, the protocol for the pyrolysis of the printed samples was established. Pyrolysis was carried out in a Nabertherm LT furnace. In a first step, the samples were heated from room temperature to 80 °C at 0.2 °C/min. The temperature was held for 5 h to dry the structure. The furnace was further heated to 200 °C at 0.4 °C/min and held for 4h to remove the physically and chemically bound water. In a final step, the system was heated to 450 °C at 0.21 °C/min and held for 4 h to pyrolyze the polymerized HDDA and remaining organics. Cylinder-shaped parts obtained after drying and pyrolysis of a structure printed with a 45 vol% $H_2O$ emulsion (2.5 mm height x 2.5 mm diameter) were heat treated in a dilatometer (DIL 806, TA Instruments) to investigate the sintering profile of the porous green body. The system was heated to 1100 °C or 1200 °C during 4 h and remained at the given sintering temperature for another 4 h, followed by another 4 h of cooling back to room temperature. On the basis of the dilatometry results, pyrolyzed samples were also sintered in a Nabertherm LT furnace in order to produce fully inorganic structures featuring a wide range of pore morphologies and final porosities.

<u>Adsorption and UV-Visible measurements</u>

Octet lattices with 4x4 unit cells and 0.375 mm strut thickness were printed from emulsions containing_0 vol% and 44.9 vol% water (10.4 mm x 10.4 mm x 10.4 mm). Bulk samples were prepared by curing the water-free resin with a UV lamp (Omnicure S1000, Lumen Dynamics). All samples were calcined according to the procedure described above and sintered at 1100 °C for 2 h to generate open porosity in the hierarchically porous ceramic. The sample masses after sintering were 88.8 mg for the porous octet, 187.1 mg for the dense octet and 141 mg for the bulk ceramic. To evaluate the ability of the printed structures to remove MB from water, the sintered ceramics were first immersed in 5 mL of an aqueous solution with 10 mg/L methylene blue (MB, >98.5% purity, Tokyo Chemical Industry) for 24 h. Fixed aliquots of 50 μL were taken from the aqueous solution at the following time points: 5 min, 10 min, 30 min, 1 h, 2 h, 4 h, 6 h and 24 h. After dilution with water by a factor 20, the solutions were characterized by UV-Vis spectroscopy using a Jasco V660 Spectrophotometer at a scan rate of 200 nm/min. The MB adsorption by the different ceramic structures was calculated based on the UV-Vis absorption at 664 nm. Pseudo-second order adsorption fitting if the experimental data was performed as described in the Supplementary Information.

Printing of graded objects

Objects with graded porosity were printed from emulsion inks with varying $H_2O$ concentrations (0 vol%, 16.9 vol%, 28.9 vol%, 44.9 vol%, 50.4 vol%) prepared following the procedure described above. A gradient in porosity along the height of the cylinder was created by changing the ink composition every 8-13 printed layers. Starting with a water-free ink, the printing process was periodically stopped to replace the resin by the next ink with higher water content until the composition with 50.4 vol% water was used. The cylinders were sintered without cracking using the protocols explained above.

SEM sample preparation and imaging

Samples for scanning electron microscopy imaging were mounted on a carbon sticker and sputtered with 5 - 7 nm Pt to improve sample conductivity. Sputter-coated samples were observed at 3 - 5 kV on a Zeiss LSM 510 electron microscope with an in-lens detector.

## References

1. Studart AR, Studer J, Xu L, Yoon K, Shum HC, Weitz DA. Hierarchical Porous Materials Made by Drying Complex Suspensions. *Langmuir* 2011, **27**(3)**:** 955-964.

2. Leys SP, Yahel G, Reidenbach MA, Tunnicliffe V, Shavit U, Reiswig HM. The Sponge Pump: The Role of Current Induced Flow in the Design of the Sponge Body Plan. *Plos One* 2011, **6**(12).

3. Fratzl P, Weinkamer R. Nature's hierarchical materials. *Prog Mater Sci* 2007, **52**(8)**:** 1263-1334.

4. Burger EH, Klein-Nulend J. Mechanotransduction in bone - role of the lacuno-canalicular network. *Faseb J* 1999, **13:** S101-S112.

5. Reznikov N, Shahar R, Weiner S. Bone hierarchical structure in three dimensions. *Acta Biomater* 2014, **10**(9)**:** 3815-3826.

6. Lakes R. Materials with Structural Hierarchy. *Nature* 1993, **361**(6412)**:** 511-515.

7. Moore DG, Barbera L, Masania K, Studart AR. Three-dimensional printing of multicomponent glasses using phase-separating resins. *Nat Mater* 2020, **19**(2)**:** 212-+.

8. Meza LR, Zelhofer AJ, Clarke N, Mateos AJ, Kochmann DM, Greer JR. Resilient 3D hierarchical architected metamaterials. *P Natl Acad Sci USA* 2015, **112**(37)**:** 11502-11507.

9. Sommer MR, Alison L, Minas C, Tervoort E, Ruhs PA, Studart AR. 3D printing of concentrated emulsions into multiphase biocompatible soft materials. *Soft Matter* 2017, **13**(9)**:** 1794-1803.

10. Alison L, Menasce S, Bouville F, Tervoort E, Mattich I, Ofner A*, et al.* 3D printing of sacrificial templates into hierarchical porous materials. *Sci Rep-Uk* 2019, **9**.

11. Muth JT, Vogt DM, Truby RL, Menguc Y, Kolesky DB, Wood RJ*, et al.* Embedded 3D Printing of Strain Sensors within Highly Stretchable Elastomers. *Adv Mater* 2014, **26**(36)**:** 6307-6312.

12. Minas C, Carnelli D, Tervoort E, Studart AR. 3D Printing of Emulsions and Foams into Hierarchical Porous Ceramics. *Adv Mater* 2016, **28**(45)**:** 9993-9999.

13. Maurath J, Willenbacher N. 3D printing of open-porous cellular ceramics with high specific strength. *J Eur Ceram Soc* 2017, **37**(15)**:** 4833-4842.

14. Schaedler TA, Jacobsen AJ, Torrents A, Sorensen AE, Lian J, Greer JR*, et al.* Ultralight Metallic Microlattices. *Science* 2011, **334**(6058)**:** 962-965.

15. Gu XW, Greer JR. Ultra-strong architected Cu meso-lattices. *Extreme Mech Lett* 2015, **2:** 7-14.

16. Portela CM, Vidyasagar A, Krodel S, Weissenbach T, Yee DW, Greer JR*, et al.* Extreme mechanical resilience of self-assembled nanolabyrinthine materials. *P Natl Acad Sci USA* 2020, **117**(11)**:** 5686-5693.

17. Zheng XY, Lee H, Weisgraber TH, Shusteff M, DeOtte J, Duoss EB*, et al.* Ultralight, Ultrastiff Mechanical Metamaterials. *Science* 2014, **344**(6190)**:** 1373-1377.

18. Susec M, Ligon SC, Stampfl J, Liska R, Krajnc P. Hierarchically Porous Materials from Layer-by-Layer Photopolymerization of High Internal Phase Emulsions. *Macromol Rapid Comm* 2013, **34**(11)**:** 938-943.

19. Cooperstein I, Layani M, Magdassi S. 3D printing of porous structures by UV-curable O/W emulsion for fabrication of conductive objects. *J Mater Chem C* 2015, **3**(9)**:** 2040-2044.

20. Binks BP, Lumsdon SO. Influence of particle wettability on the type and stability of surfactant-free emulsions. *Langmuir* 2000, **16**(23)**:** 8622-8631.

21. Aveyard R, Binks BP, Clint JH. Emulsions stabilised solely by colloidal particles. *Adv Colloid Interfac* 2003, **100:** 503-546.

22. Mason TG. New fundamental concepts in emulsion rheology. *Curr Opin Colloid In* 1999, **4**(3)**:** 231-238.

23. Wang KY, Pan WY, Liu Z, Wallin TJ, van Dover G, Li S*, et al.* 3D Printing of Viscoelastic Suspensions via Digital Light Synthesis for Tough Nanoparticle-Elastomer Composites. *Adv Mater* 2020, **32**(25).

24. Jacobs PFaR, D.T. and Computer and Automated Systems Association of SME. *Rapid Prototyping & Manufacturing: Fundamentals of Stereolithography*. Society of Manufacturing Engineers, 1992.

25. Schwentenwein M, Homa J. Additive Manufacturing of Dense Alumina Ceramics. *Int J Appl Ceram Tec* 2015, **12**(1)**:** 1-7.

26. Meza LR, Phlipot GP, Portela CM, Maggi A, Montemayor LC, Comella A*, et al.* Reexamining the mechanical property space of three-dimensional lattice architectures. *Acta Mater* 2017, **140:** 424-432.

27. Banerjee S. On the mechanical properties of hierarchical lattices. *Mech Mater* 2014, **72:** 19-32.

28. Gibson LJ, Ashby MF. *Cellular solids structure and properties*, 2nd edn. Cambridge University Press: Cambridge, 1997.

29. Schaedler TA, Carter WB. Architected Cellular Materials. *Annu Rev Mater Res* 2016, **46:** 187-210.

30. Ashby MF. The properties of foams and lattices. *Philos T R Soc A* 2006, **364**(1838)**:** 15-30.

31. Portela CM, Greer JR, Kochmann DM. Impact of node geometry on the effective stiffness of non-slender three-dimensional truss lattice architectures. *Extreme Mech Lett* 2018, **22:** 138-148.

32. Hutchinson RG, Fleck NA. Microarchitectured cellular solids - the hunt for statically determinate periodic trusses - Plenary lecture presented at the 75th Annual GAMM

Conference, Dresden, Germany, 22-26 March 2004. *Zamm-Z Angew Math Me* 2005, **85**(9)**:** 607-617.

33. Vigliotti A, Pasini D. Mechanical properties of hierarchical lattices. *Mech Mater* 2013, **62:** 32-43.

34. Ani SM, Muchtar A, Muhamad N, Ghani JA. Binder removal via a two-stage debinding process for ceramic injection molding parts. *Ceram Int* 2014, **40**(2)**:** 2819-2824.

35. Ho YS, Ofomaja AE. Kinetic studies of copper ion adsorption on palm kernel fibre. *J Hazard Mater* 2006, **137**(3)**:** 1796-1802.

36. Aizenberg J, Weaver JC, Thanawala MS, Sundar VC, Morse DE, Fratzl P. Skeleton of Euplectella sp.: Structural hierarchy from the nanoscale to the macroscale. *Science* 2005, **309**(5732)**:** 275-278.

37. Weaver JC, Aizenberg J, Fantner GE, Kisailus D, Woesz A, Allen P*, et al.* Hierarchical assembly of the siliceous skeletal lattice of the hexactinellid sponge Euplectella aspergillum. *J Struct Biol* 2007, **158**(1)**:** 93-106.

38. Weaver JC, Milliron GW, Allen P, Miserez A, Rawal A, Garay J*, et al.* Unifying Design Strategies in Demosponge and Hexactinellid Skeletal Systems. *J Adhesion* 2010, **86**(1)**:** 72-95.

39. Chen GD, Zhou SX, Gu GX, Yang HH, Wu LM. Effects of surface properties of colloidal silica particles on redispersibility and properties of acrylic-based polyurethane/silica composites. *J Colloid Interf Sci* 2005, **281**(2)**:** 339-350.

40. Lee JH, Prud'homme RK, Aksay IA. Cure depth in photopolymerization: Experiments and theory. *J Mater Res* 2001, **16**(12)**:** 3536-3544.

41. Tomeckova V, Halloran JW. Cure depth for photopolymerization of ceramic suspensions. *J Eur Ceram Soc* 2010, **30**(15)**:** 3023-3033.

42. Tomeckova V, Halloran JW. Critical energy for photopolymerization of ceramic suspensions in acrylate monomers. *J Eur Ceram Soc* 2010, **30**(16)**:** 3273-3282.

43. Wypych G. *Handbook of material weathering*, Sixth edition edn, 2018.

## Table of Contents

**Three-dimensional printing of particle-stabilized emulsions enables hierarchical porous composites and ceramics with unprecedented shape complexity**

Photo-curable, particle-stabilized emulsions can be used to generate hierarchical porous composites via self-assembly of particles and droplets at the microscale and layer-by-layer light-induced polymerization at the macroscale. Sintering of the as-printed composites results in hierarchical porous ceramics with exquisite porosity and shape control at three hierarchical levels.

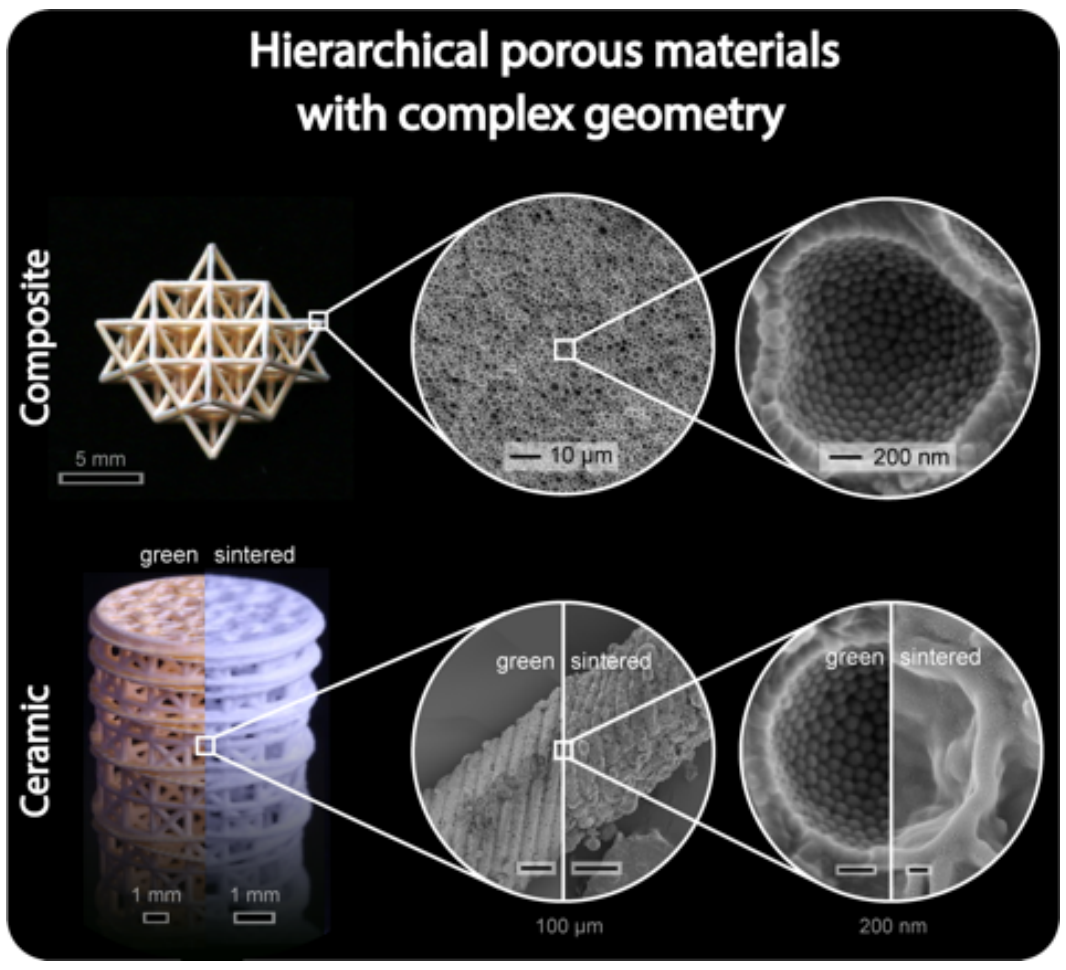

# Supplementary Information: Hierarchical porous materials made by stereolithographic printing of photo-curable emulsions

Nicole Kleger,[1] Clara Minas,[1] Patrick Bosshard,[1] Iacopo Mattich[1,2], Kunal Masania,[1,2]* André R. Studart [1]*

[1] Complex Materials, Department of Materials, ETH Zurich, 8093 Zurich, Switzerland

[2] Soft Materials, Department of Materials, ETH Zurich, 8093 Zurich, Switzerland

[3] Current address: Faculty of Aerospace Engineering, Delft University of Technology, Kluyverweg 1, 2629 HS Delft, Netherlands

*corresponding authors: k.masania@tudelft.nl, andre.studart@mat.ethz.ch

**Content:**

**Influence of monomer ratio and nanoparticle content on the droplet size**

In addition to the $H_2O$ content, the PUA monomer and the $SiO_2$ particles may also affect the type of emulsion formed and the droplet size achieved. Varying the concentration of PUA relative to the other monomer (HDDA) in the emulsion influences significantly the pore size distribution of the printed structures (Supplementary Figure 1). For emulsions with a fixed water content of 45 vol% (40 wt%), we found that the average pore size decreases from 1.97 ± 1.54 µm to 0.50 ± 0.26 µm by increasing the PUA concentration from 0 to 10 wt% relative to the mass of HDDA. Such decrease in average droplet size is accompanied by a significant narrowing of the pore size distribution. When the PUA concentration is further increased above 28 wt%, a transitional phase inversion from a water-in-oil to an oil-in-water emulsion is observed. Such transition suggests that the PUA monomer reduces the wettability of the modified silica particles in the oil phase. This favours the displacement of the interfacially-adsorbed particles towards the water phase, which can eventually cause the emulsion to phase invert at sufficiently high PUA concentrations. This interpretation is in line with previous reports in the literature[21, 39] and provides a rationale for the monomer composition selected to tailor the wetting and self-assembly behaviour of the particles within the emulsion.

To investigate the possible effect of the silica concentration on the emulsion structure, we prepared formulations with 10.9 – 27.1 wt% $SiO_2$ and fixed PUA and $H_2O$ concentrations of 10 wt% (with respect to HDDA) and 40 wt% (44.9 vol%), respectively. Increasing the $SiO_2$ concentration was found to improve the monodispersity of the pore size, while only slightly decreasing the mean droplet and pore sizes (Supplementary Figure 2). This is in line with an expected coalescence of droplets due to reduced particle surface coverage of the water-oil interface at lower silica concentrations. The initially < 500 nm droplets generated under tip sonication are expected to show a particle surface coverage below a monolayer (< 90%) if the silica nanoparticle concentration is lower than 20 wt%. The incomplete surface coverage will hence inevitably result in droplet coalescence and hence broader size distribution.

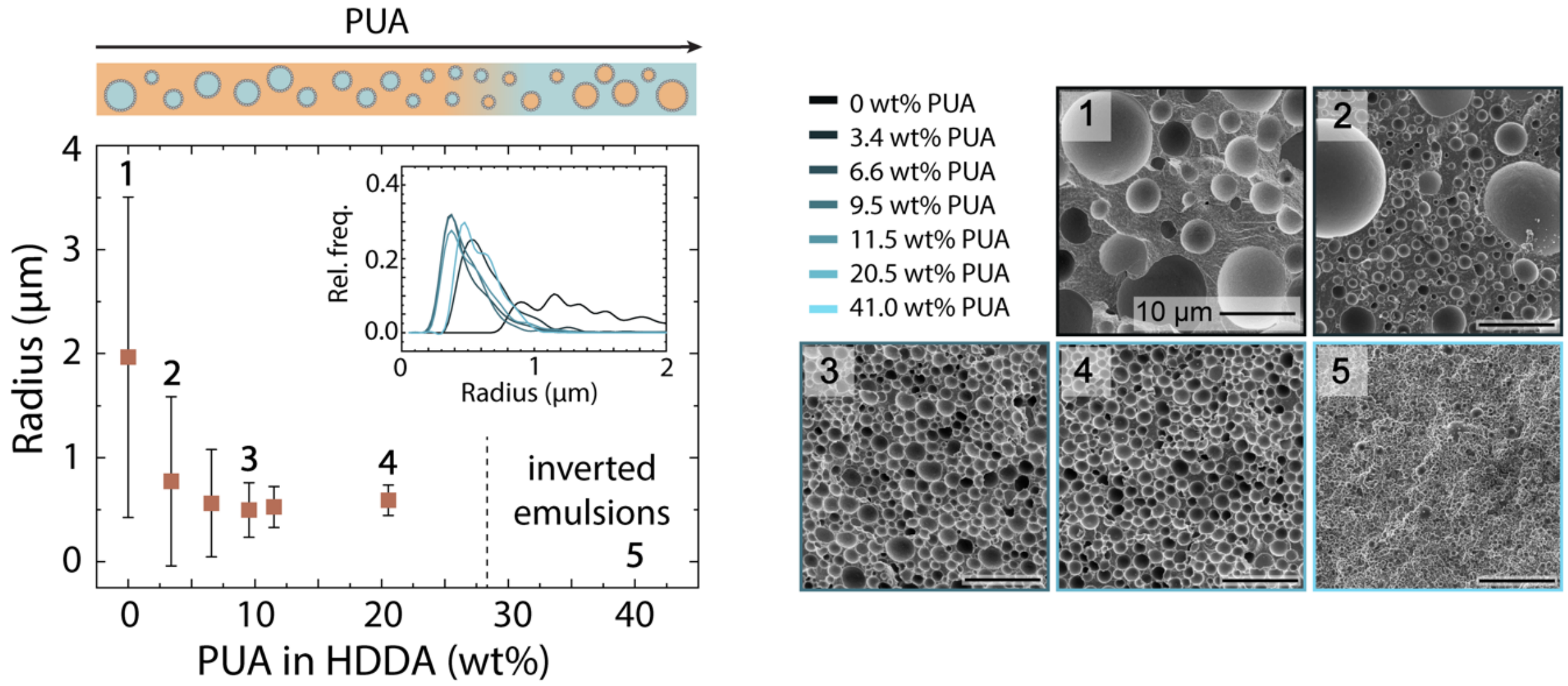

**Supplementary Figure 1.** Effect of the concentration of the PUA monomer on the type of emulsion and the size of droplets achieved upon mixing (left) and on the structure of the cured resin after drying and curing (right). The corresponding droplet size distributions are shown as an inset. Scale bar: 10 µm.

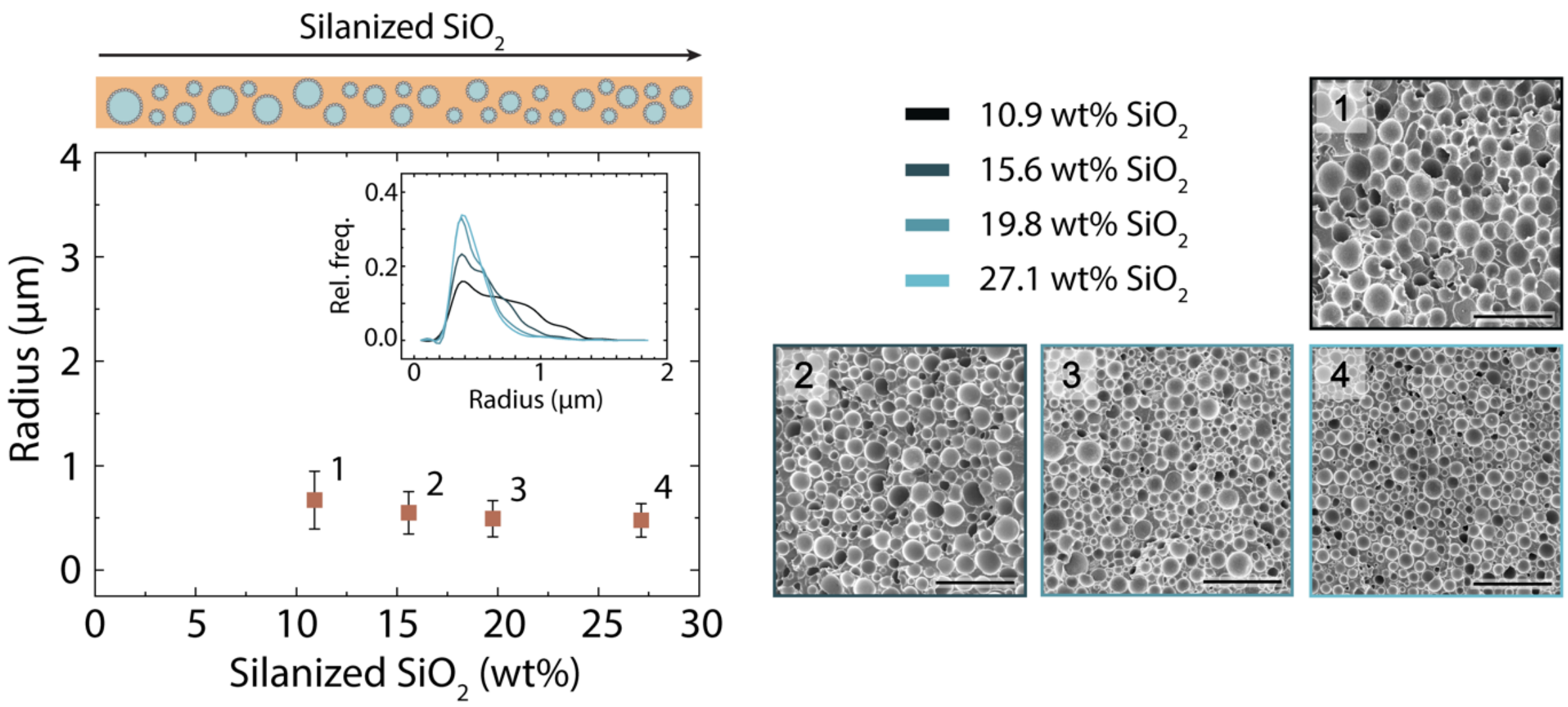

**Supplementary Figure 2.** Influence of the concentration of the $SiO_2$ nanoparticles on the size of droplets achieved upon emulsification (left) and on the structure of the cured resin after drying (right). The corresponding distributions of the droplet sizes are shown as an inset. Scale bar: 10 µm.

**Rheological behavior of Pickering emulsion resins**

Full rheological characterization of emulsion resins without (Supplementary Figure 3a) or with 10 wt% PUA (Supplementary Figure 3b) was performed for water contents of 0 vol%, 29 vol% or 45 vol%. All measurements were taken on an Anton Paar MCR 302 rheometer (Anton Paar GmbH) using a setup comprising of a sand-blasted 25 mm parallel plate positioned at a 1 mm gap size. Amplitude sweeps were performed at logarithmically increasing shear stresses from 0.0001 to 100 Pa at a constant frequency of 1 rad/s to obtain the storage modulus *G'* and loss modulus G'' (Supplementary Figure 3, I). Additionally, strain-controlled frequency sweeps (Supplementary Figure 3, II) were performed by logarithmically increasing the angular frequency from 0.1 to 500 rad/s at a constant shear strain of 0.01%. The resulting *G''* values were used to calculate the dynamic viscosity of the emulsions (Supplementary Figure 3d), which is given by $G''/\omega$, where $\omega$ denotes the angular frequency. The reported high-frequency viscosities were calculated for angular frequencies higher than 100 rad/s, above which the dynamic viscosity was found to be stable. Finally, steady-state measurements were performed by logarithmically increasing shear stresses from 0.01 to 100 Pa. The yield stress (Supplementary Figure 3e) was taken as the point at which the shear strain suddenly increases in the strain-stress curves obtained from the steady-state measurements. All measurements were performed at 25 °C.

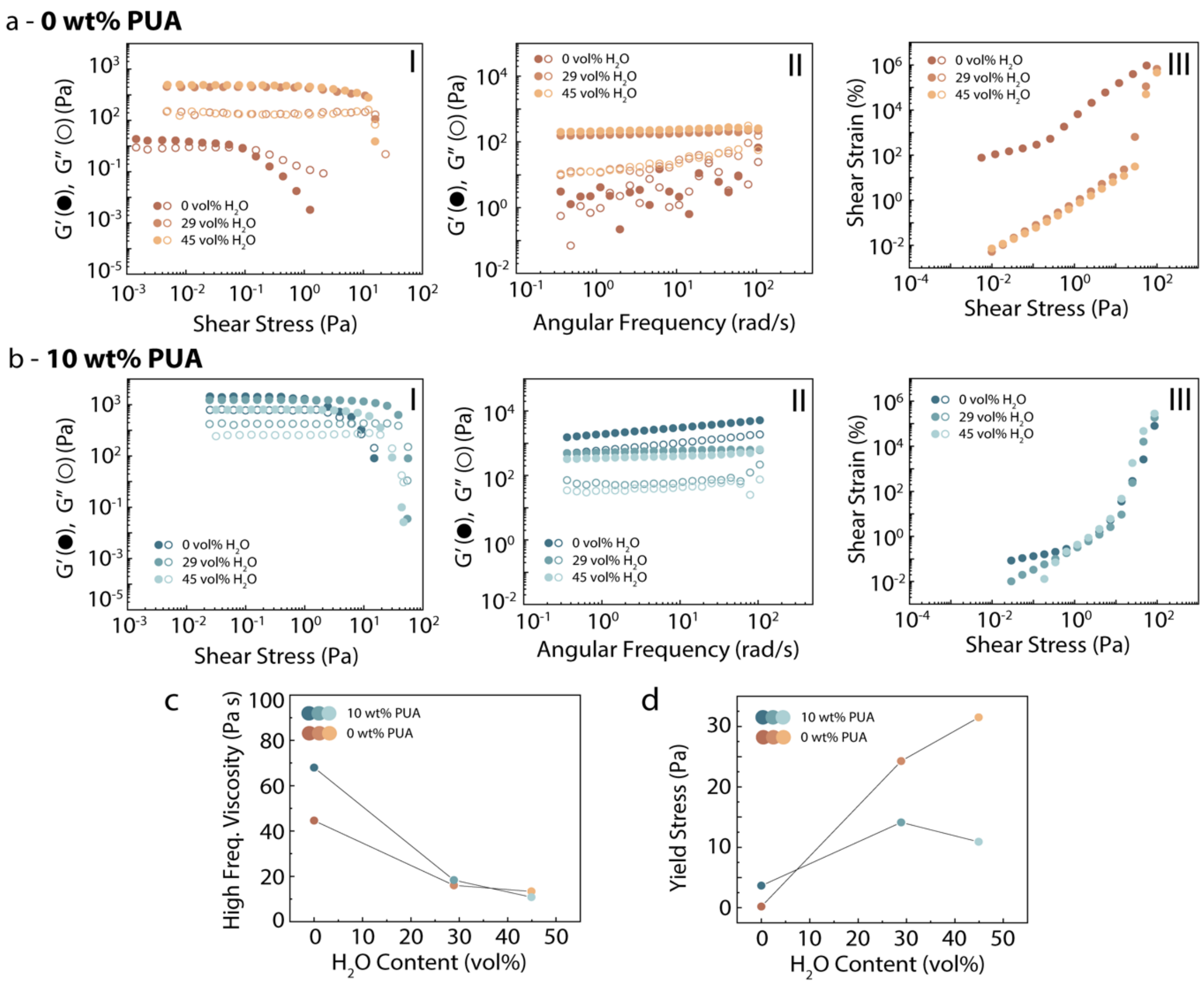

**Supplementary Figure 3.** Rheological behavior of emulsions prepared with water contents of 0%, 25% and 50% vol, (a) with and (b) without the addition of PUA monomer, consisting of an amplitude sweep (I), frequency sweep (II) and steady state measurement (III). (c) High-frequency dynamic viscosity and (d) yield stress of emulsions with and without PUA as a function of the water content.

**Polymerization behaviour of Pickering emulsion resins**

To identify emulsion formulations and printing conditions that lead to high-fidelity porous structures, we first analyse the effect of compositional and processing parameters on the polymerization behaviour of the reactive layer. Using Lambert-Beer's law to quantify the decay in light intensity along the build direction, the effect of the initiator and UV-blocker concentrations on the measured thickness of the polymerized layer can be predicted reasonably well, allowing us to gain insights into the physical mechanisms underlying the polymerization process (Figure 2a,b, main text).

According to the Lambert-Beer mode, the thickness of the polymerized layer ($z_p$) depends on the light dose ($D$) as follows[24]: $z_p = h_a * \ln\left(\frac{D}{D_c}\right)$, where $h_a$ is the light penetration depth and $D_c$ is the critical dose of light required to initiate polymerization. Single-layer printing experiments were performed to verify the validity of this analytical expression in predicting the polymerization behaviour of photo-curable emulsions with selected concentrations of initiator and UV-blocker (Figure 2a,b). In these experiments, the light dose $D$ (units of energy) is deliberately varied by changing the exposure time at a constant light intensity (units of power).

The results show that the theoretical model predicts the effect of the initiator and UV-blocker concentrations on the measured thickness of the polymerized layer reasonably well, allowing us to gain insights into the physical mechanisms underlying the polymerization process. By plotting the measured thickness as a function of the log($D$), we can directly obtain the penetration depth $h_a$ and the critical dose $D_c$ from the slope and the x-axis intercept of the theoretical fits, respectively. For compositions with fixed UV-blocker content, the analysis reveals that an increase in initiator concentration initially reduces the critical dose necessary for polymerization ($D_c$) while keeping the penetration depth ($h_a$) mostly unchanged (Figure 2a). If the nominal initiator concentration surpasses a threshold value of 0.5wt% (with respect to the monomer), no effect on the critical light dose is observed. These findings are in line with earlier reports[40, 41, 42] and reflect the fact that a minimum concentration of initiator molecules need to be photo-activated to initiate the chain growth reaction. Our data suggest that this minimum amount of reactive molecules is reached for formulations with a nominal initiator content of at least 0.5 wt% combined with a light dose equal or higher than 3 $mJ/cm^2$. Below this nominal initiator content, the number of reactive species can be increased by enhancing the light dose applied, which explains the observed correlation between $D_c$ and the amount of initiator at low nominal concentrations. Conversely, higher UV-blocker concentrations were found to decrease the penetration depth ($h_a$) without affecting so much the critical light dose $D_c$ (Figure 2b). This dependence qualitatively follows the inverse correlation between penetration depth and the concentration of light absorbing species expected from Lambert-Beer's law.[43]

**Stereolithographic printing of high-fidelity structures**

Understanding the polymerization behaviour of the photo-curable emulsions enables the design of formulations suitable for printing cellular materials with programmable pore size and strut thickness, while minimizing the required light dose and printing time. We quantify the fidelity and the programmability of the stereolithographic process by printing specimens featuring large arrays of holes or pillars with pre-defined sizes along the build direction or within the plane of the substrate (x-y plane), respectively. Holes and pillars are used in these experiments as simplified negative and positive features analogous to pores and struts in a printed structure. These structures were printed with a constant light intensity of 20 mW/cm$^2$ at varying illumination times of 2, 3 or 4 s per layer, resulting in doses of 40, 60 and 80 mJ/cm$^2$, respectively. Experiments were carried out using an emulsion formulation containing the threshold initiator concentration of 0.5wt% and an UV-blocker content (0.075 wt%) that should lead to a nominal printed layer thickness of 50 µm.

Although light doses as low as 3 mJ/cm$^2$ are enough to create a gelled polymerized layer, we experimentally observed that fully cured layers require doses of at least 40 mJ/cm$^2$. For an applied light dose of 40 mJ/cm$^2$, the size of holes printed along the build direction ($d_{exp}$) were found to scale linearly with the input target values ($d_{tar}$) with a slight offset $\delta$ (Supplementary Figure 3a). The offset might be attributed to local over-polymerization of the resin caused by the diffusion of initiator molecules into non-exposed areas. Such undesired effect is probably the reason for the inaccuracy observed for holes smaller than 500 µm and printed at the higher doses of 60 and 80 mJ/cm$^2$. In this case, over-polymerization leads to holes with oval shapes that are much smaller than the target dimensions ($d_{tar}$).

For the experiments with pillars, we found that the pillar size ($d_{exp}$) scales linearly with the target value if $d_{tar}$ is at least 4 times larger than the thickness of a single polymerized layer (>200 µm, Supplementary Figure 3b). High light doses of 60 and 80 mJ/cm$^2$ lead to over-polymerization and thicker dimensions compared to the target values, whereas thinner pillar sizes are obtained for the lowest light dose (40 mJ/cm$^2$). Below the critical size of 200 µm, a poor correlation between experimental and target values is observed, likely due to the limited lateral resolution of the desktop printer used in the study.

Overall, these model experiments indicate that pore sizes and strut thicknesses down to 200 µm can be printed with high accuracy if the measured offsets between the experimental and target values (Figure 2c and Supplementary Figure 3) are taken into account in the geometrical design.

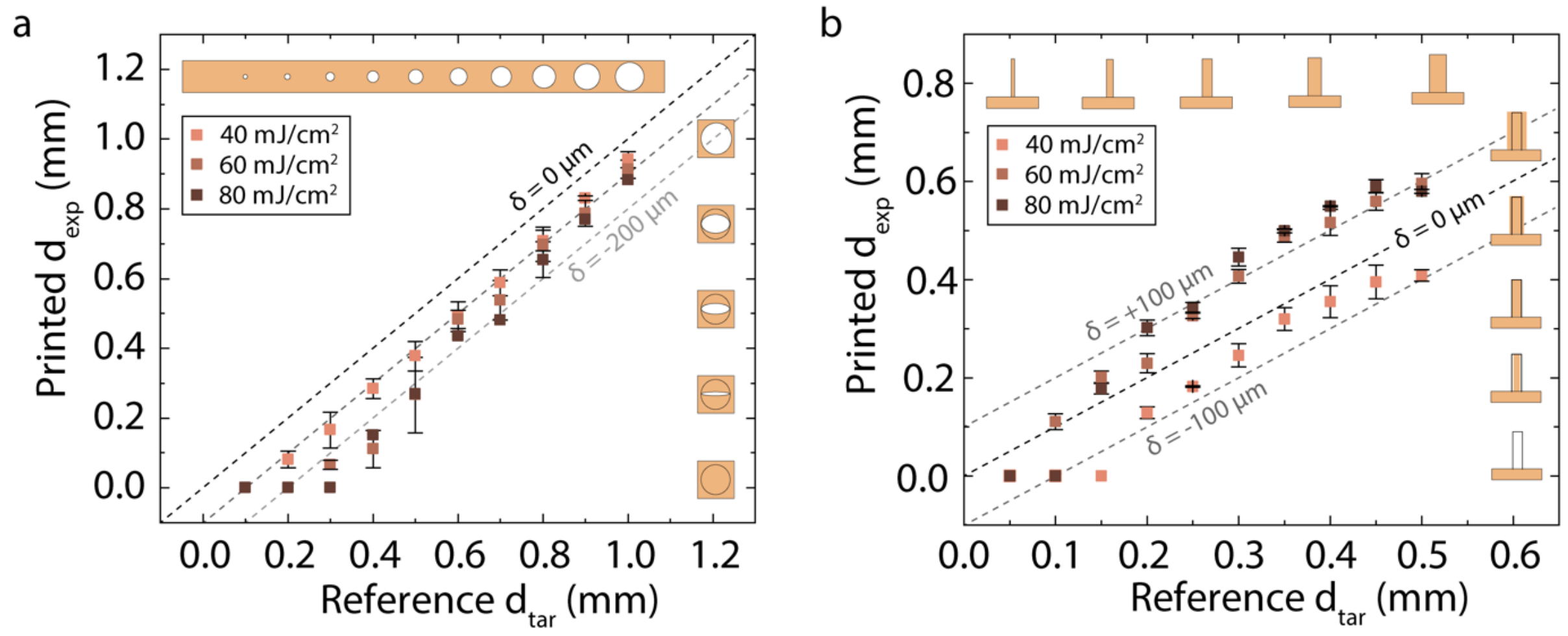

**Supplementary Figure 4.** Printing fidelity achieved for negative (holes) and positive (pillar) features using different illumination doses. Dose levels of 40, 60 and 80 mJ/cm$^2$ were applied by using exposure times of 2, 3 and 4 s, respectively, at a fixed light intensity of 20 mW/cm$^2$. The dashed line represents the perfect print, corresponding to a 1:1 translation of the print file to the printed object. Different values for the offset $\delta$ are included in the plots.

**Elastic modulus of hierarchically porous lattices with tunable strut microporosity**

In addition to changes in the strut thickness (main text), a second strategy to vary the relative density of hierarchical lattices consists in the incorporation of an increasing volume fraction of micropores in the struts up to 50 vol% while keeping their thickness constant at 0.5 mm (Supplementary Figure 4). Analogous to hierarchical structures with varying strut thickness, this approach enhances the mechanical efficiency of bending-dominated Kelvin lattices and promotes a stronger drop in elastic modulus of stretching-dominated octet lattices. The scaling exponent decreases from 2.77 to 1.48 for Kelvin structures and increases from 1.88 to 2.54 for octet lattices. These shifts in power exponents can be rationalized following the same arguments used to explain the behaviour of hierarchical structures displaying different strut thicknesses.

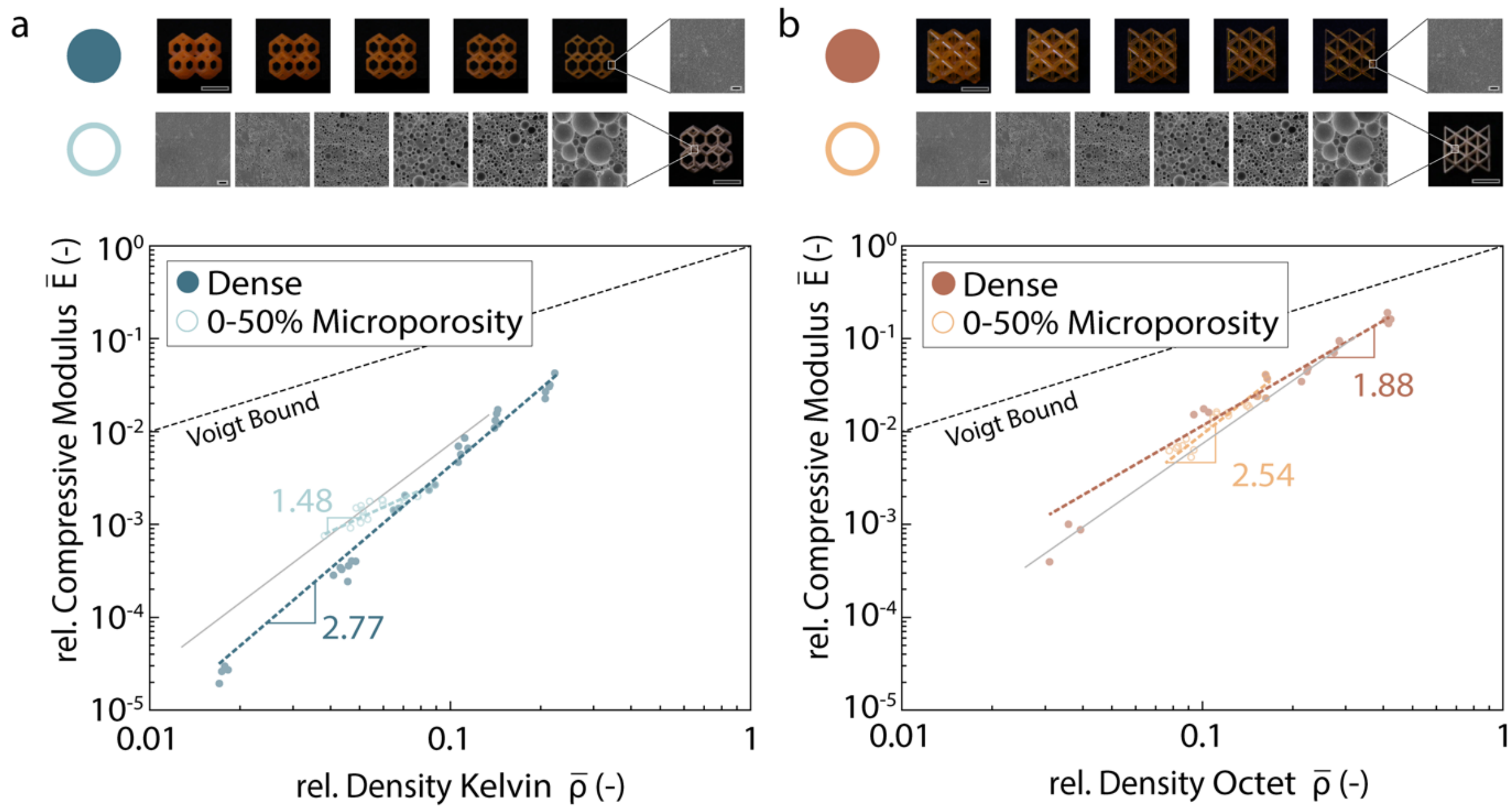

**Supplementary Figure 5.** Relative compressive modulus of (a) Kelvin and (b) octet lattices with dense struts of variable thickness (dense) compared to counterparts with fixed strut thickness (0.5 mm) and microporosity varies within the range of 0 – 50 vol% inside the struts. The coloured dashed lines show fittings obtained using the analytical model. The numbers next to such lines correspond to the power exponents. The Voigt bound is displayed as a black dashed line, whereas the full grey lines shown in (a) and (b) indicate the fittings obtained for the lattices with 45 vol% porous struts of varying strut thickness.

**Maximum compressive strength of hierarchically porous lattices**

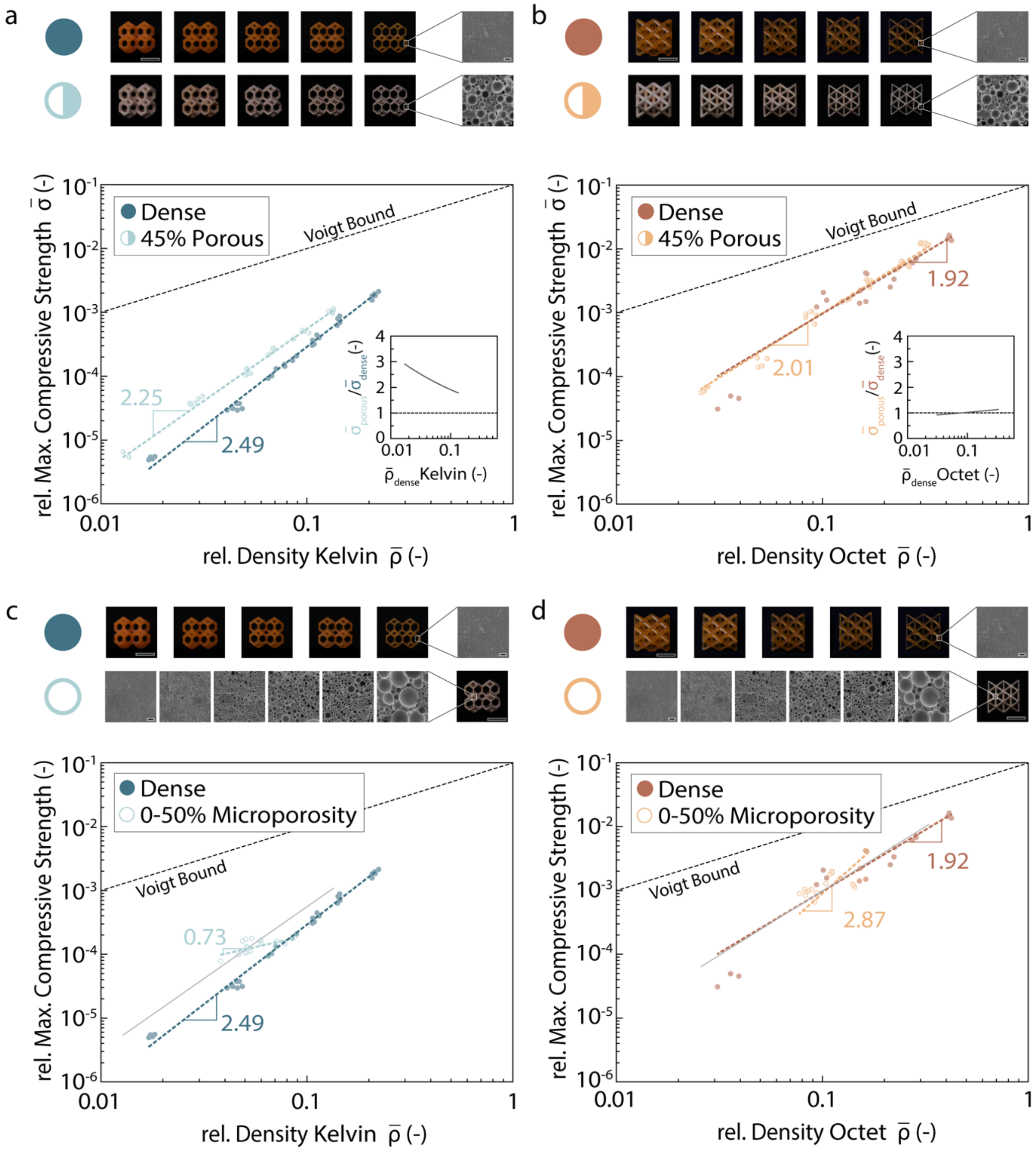

**Supplementary Figure 6.** Mechanical strength of Kelvin and octet hierarchical lattices with varying relative densities. (a,b) Relative compressive strength ($\bar{\sigma}$) of (a) Kelvin and (b) octet lattices with struts that are either dense or 45% porous. In these samples the relative density is changed by varying the strut thickness. The insets show the ratio between the relative maximum compressive strength for microporous and dense Kelvin lattices ($\bar{\sigma}_{porous}/\bar{\sigma}_{dense}$) at varying relative densities of the overall structure ($\bar{\rho}$). (c,d) Relative compressive strength of (c) Kelvin and (d) octet lattices with dense struts of variable thickness (dense) compared to counterparts with fixed strut thickness (0.5 mm) but microporosity varying between 0 and 50 vol% inside the struts. The coloured dashed lines show fittings obtained using the analytical model. The numbers next to such lines correspond to the power exponents. The Voigt bound is displayed as a black dashed line, whereas the full grey lines shown in (c) and (d) indicate the fittings obtained for the lattices with 45 vol% porous struts of varying thickness. Scale bars: 5 mm in the macroscopic images of lattices (Kelvin and octet); 10 μm in the SEM close-ups of dried emulsions.

**Pyrolysis and sintering of printed hierarchical structures**

As-printed composites can be converted into inorganic hierarchical structures by pyrolysis of the organic phase followed by sintering of the remaining oxide phase. The processing conditions required for slow removal of the organic phase during calcination were evaluated by performing thermogravimetric analysis (TGA) of the as-printed composites (Supplementary Figure 6). The TGA data reveals that heating of the printed material in air results in two main mass loss processes between 50 and 500 °C. At lower temperatures in the range of 25-200 °C, the material loses about 18% of its initial weight, which is associated with the evaporation of residual physically and chemically bound water from the structure. Heating the sample further up to the temperature window 300-500 °C leads to two sequential drops in weight and an additional mass loss of 58%. These processes correspond to the thermal oxidation and removal of the polymer phase, after which the structure is finally ready for sintering.

The sintering behaviour of the calcined structures was investigated by measuring the dimensional changes of 45 vol% porous specimens upon heating up to 1100 °C and 1200 °C (Supplementary Figure 6a). Shrinkage of the structure starts already at 500 °C and continues at the dwell temperatures to reach values between 10 and 42% at the end of the sintering process. Importantly, the sintering time at high temperatures strongly affects the structure of the material at the microscale. SEM images of sintered samples indicate that the assembly of silica nanoparticles initially present on the walls of the microscale pores undergo extensive morphological changes during the heat treatment (Supplementary Figure 6b). At 1100 °C, the pore wall morphology changes within the first hour of sintering from a network of particles hold by interconnecting necks to a dense assembly of grains around closed pores. For longer sintering times, the spherical pores eventually transform into an open porous network with coral-like morphology. When sintered at the higher temperature of 1200 °C, this open network finally coalesces into a denser microstructure featuring isolated closed pores. The rich morphologies generated upon sintering allows one to tune the thermal treatment to develop open or closed microporosity depending on the target application.

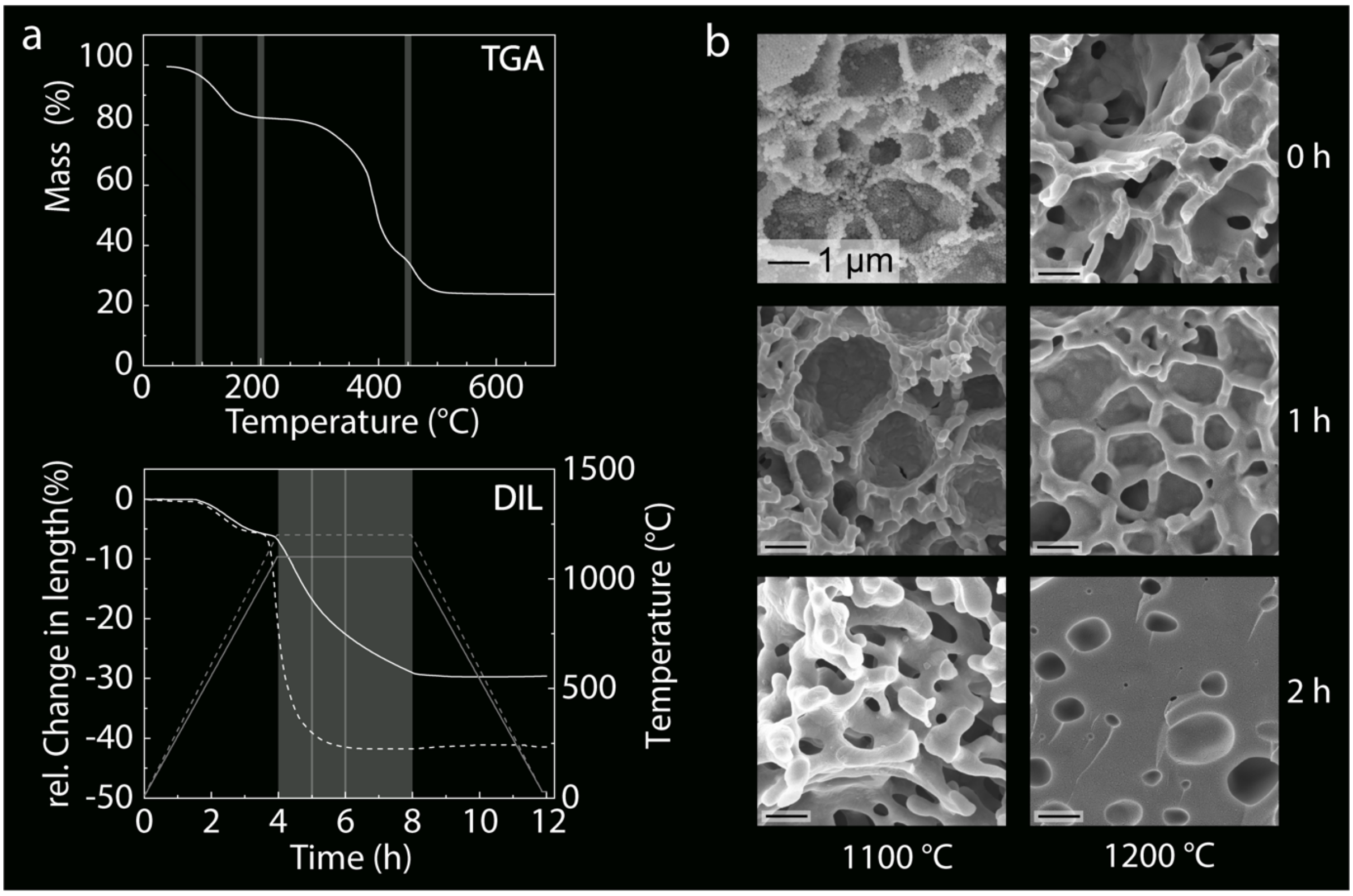

**Supplementary Figure 7.** Conversion of as-printed composites into inorganic hierarchical structures. (a) Thermogravimetric analysis (TGA) and dilatometry (DIL) of the as-printed composite, indicating the temperatures at which the organic phase is removed from the sample during calcination and the shrinkage of the material during sintering at high temperatures, respectively. The grey vertical lines in the TGA graph indicate the hold temperatures used for the calcination process. The shrinkage of calcined samples during sintering in shown in the DIL plot for heat treatments at 1100°C (full line) and 1200°C (dashed line). (b) SEM images depicting the microstructure of the struts after sintering at 1100 or 1200 °C for different elapsed times.

**SLA printed object with graded porosity**

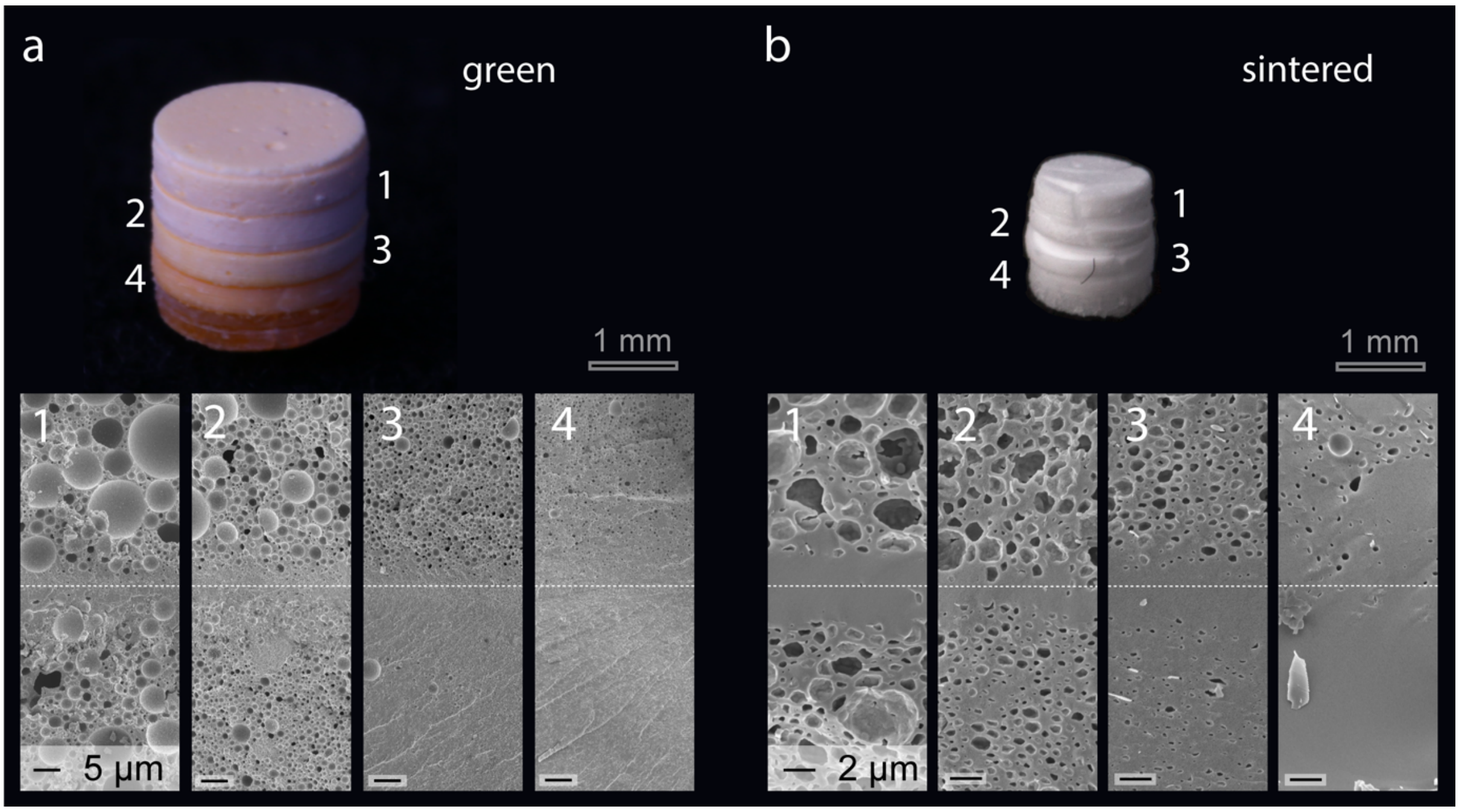

**Supplementary Figure 8.** Cylindrical-shaped objects with graded porosity before and after sintering. (a) As-printed (green) parts with graded porosity along the height prepared by varying the emulsion water content in individual polymerized layers. The numbers 1, 2, 3 and 4 indicate boundaries of the layers obtained using water contents in the ranges of 50-45 vol%, 29-45 vol%, 17-29 vol%, and 0-17 vol%, respectively. SEM close-ups of the interfaces between layers are also shown. (b) Sintered silica part with graded porosity along the height. SEM images depicting the microstructures of the graded cylinder-shaped object after sintering at 1200°C for 1 h. The graded porosity resulting from the increasing $H_2O$ content of the emulsion was preserved during the sintering process. Scale bar in (b): 2 µm**.**

**Adsorption properties of hierarchical porous structures**

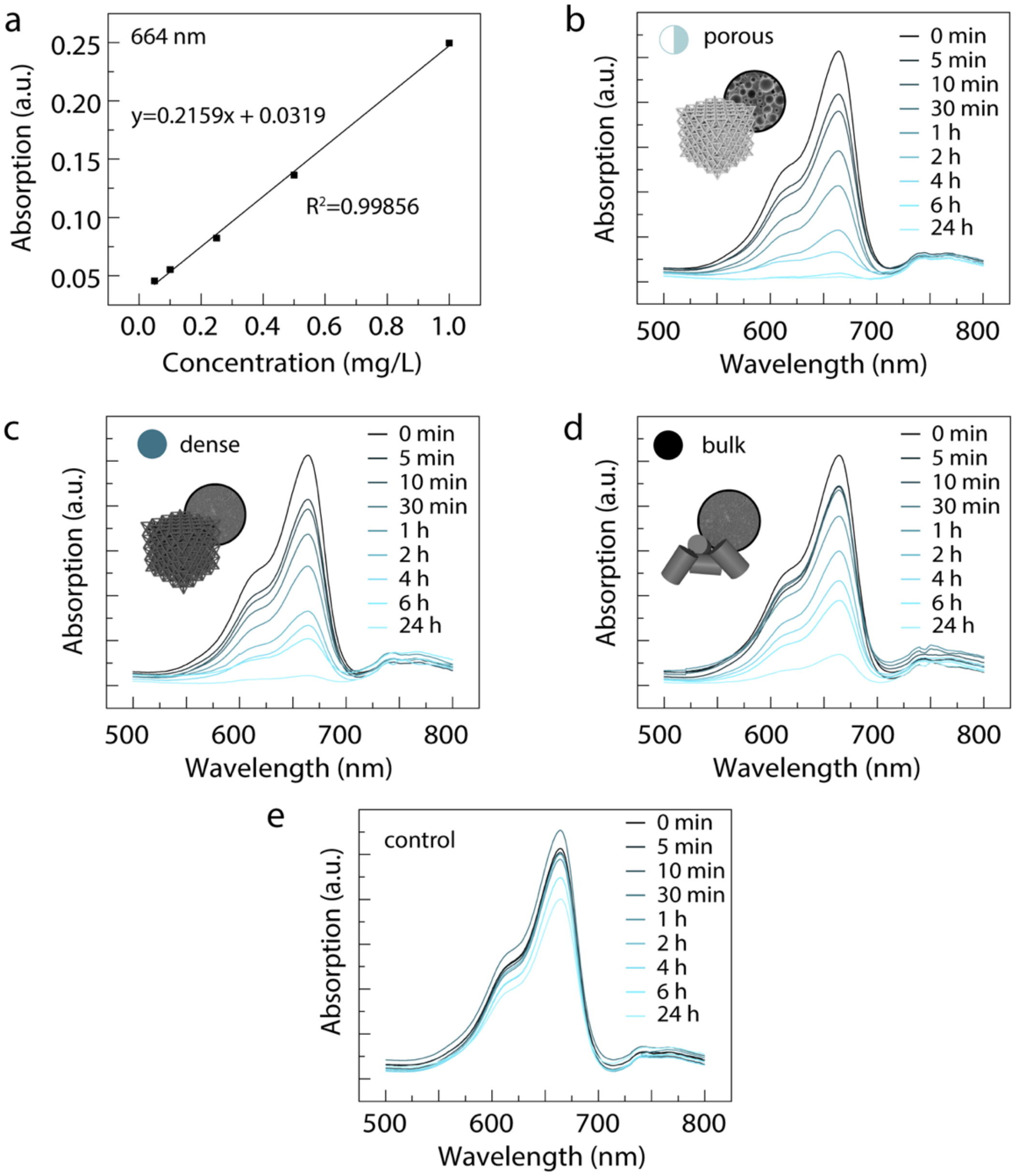

**Supplementary Figure 9.** Light absorption experiments performed to assess the adsorption performance of hierarchical porous structures (Figure 4c,d, main text). (a) Calibration curve correlating the light absorption at 664nm and the concentration of methylene blue in aqueous solution. (b-d) UV-Vis analysis of the MB aqueous solutions into which octet lattices with microporous (b) and dense struts (c) were immersed for distinct time periods. Plot (d) shows the data obtained for the reference bulk ceramic piece. 5mL of an aqueous solution with initial concentration of 10 mg/L methylene blue was used in all the experiments. The UV-Vis measurements show that the octet lattice with microporous struts has higher adsorption and is a factor eight faster than the bulk ceramic and twice as fast as the octet structure with dense struts over 24 hours of time. Control measurements without adsorbent are shown in (e).

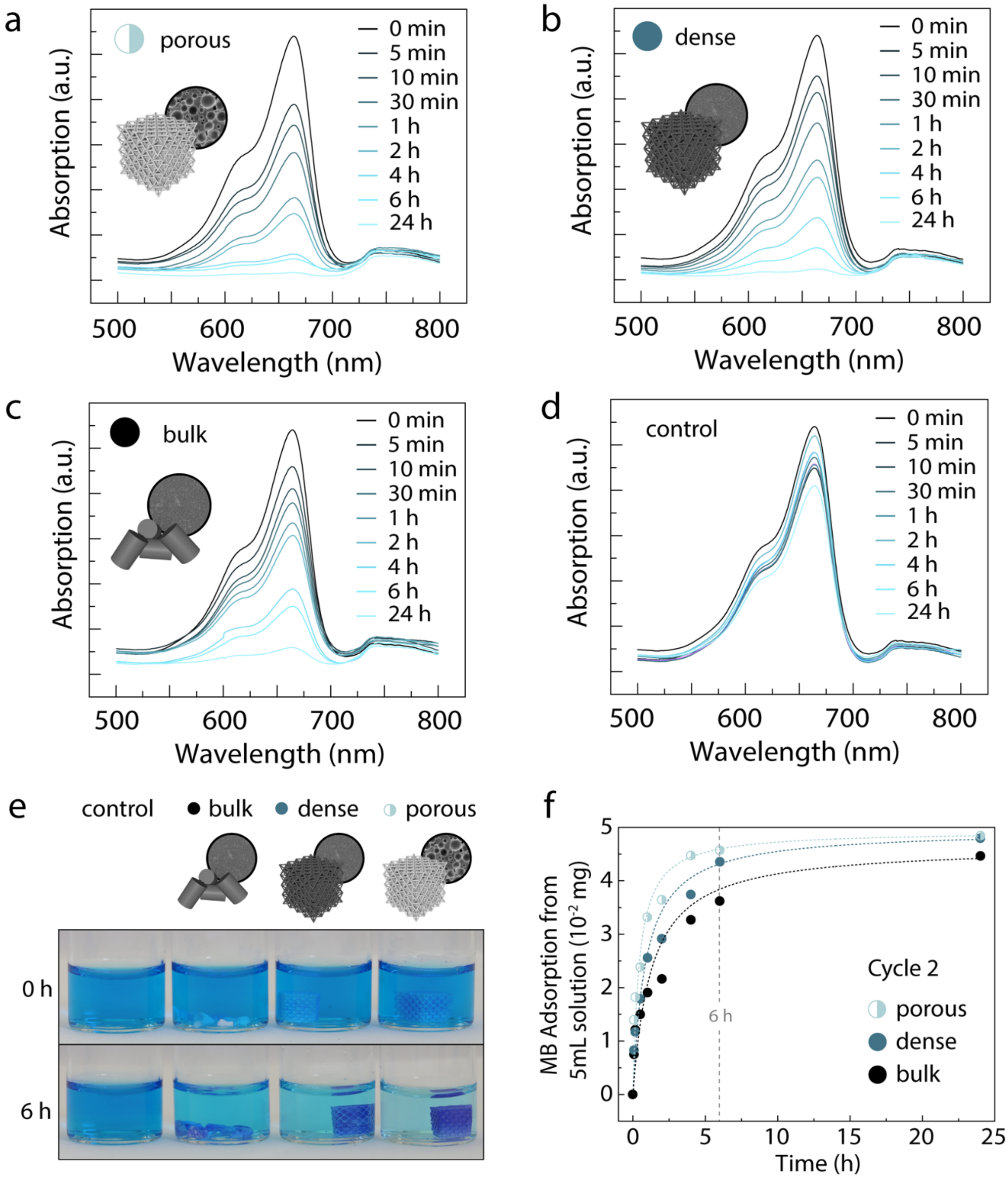

**Supplementary Figure 10.** Adsorption capacity of hierarchical porous structures that were recycled from the first experimental series (Figure 4c,d and Supplementary Figure 9). (a-c) UV-Vis analysis of the MB aqueous solutions into which octet lattices with (a) microporous and (b) dense struts were immersed for distinct time periods. A bulk ceramic piece was also tested for comparison (c). The second cycle of adsorption tests demonstrates that the structures can be recovered and reused at equal performance. Control measurements without adsorbent are shown in (e). 5 mL of an aqueous solution with initial concentration of 10 mg/L methylene blue was used in all the experiments.

**Pseudo-second order kinetic adsorption fitting of methylene blue adsorption**

The adsorption kinetics of the ceramic structures was quantified by fitting the experimental adsorption ($Q$ in mg) data with the following pseudo-second-order kinetics equation:

$$Q = \frac{Q_e k_2 t}{Q_e k_2 t + 1}$$

where $k_2$ is a fitting parameter, $Q_e$ is the equilibrium adsorption in mg and $t$ is the time of adsorption in hours.

The obtained fittings are shown as dashed lines in Figure 4d (adsorption cycle 1) and in the Supplementary Figure 10f (adsorption cycle 2).